\documentstyle[12pt,fleqn]{article}
\input epsf.sty
\begin{document}
\newcommand{\paper}[3]{\noindent\begin{tabular}
{p{6.5cm}p{1cm}p{7.5cm}}{#1}&{#2}&{#3} 
\end{tabular}}
\language=0
\begin{center}
{\large\bf Influence of wave frequency variation on anomalous
cyclotron resonance interaction of energetic electrons with
finite amplitutude ducted whistler-mode wave.}
\end{center}
\vskip0.8cm
\centerline{{\sc N.S.Erokhin$,^*$ N.N.Zolnikova$,^*$ M.J.Rycroft$\,
^{\dagger}$
and D.Nunn$\, ^{\ddagger}$}}

\begin{quote}
\par
{\small $\; ^*$Space Research Institute of the Russian Academy of Sciences,
Profsoyuznaya Str. 84/32, Moscow 117810, Russia; $^{\dagger}$International
Space Research University, Parc d'Innovation, Boulevard Genthier d'Andernach,
F-67400 Illkirch, France; $\ddagger$Department of Electronics and Computer
Sciences, University of Southampton, SO9 5NH, U.K.}
\end{quote}
\vskip1.2cm
\begin{quote}
{\small{\bf Abstract}-- The influence of wave frequency variation on the
anomalous cyclotron resonance $\omega=\omega_{Be}+kv_{\|}$ interaction
(ACRI) of energetic electrons with a ducted finite amplitude whistler-mode
wave propagating through the so-called transient plasma layer (TPL)
in the magnetosphere or in the ionosphere is studied both analytically and
numerically. The anomalous cyclotron resonance interaction takes place in
the case when the whistler-mode wave amplitude $B_{W}$ is consistent with
the gradient of magnetic field $\overrightarrow{B_0}$. The region of phase
space occupied by anomalously interacting energetic electrons (synchronous
particles) is determined. The efficiencies of both the pitch-angle scattering
of resonant electrons and their transverse acceleration are studied and the
efficiencies dependence on the magnitude and sign of the wave frequency drift
is considered. It has been shown that in the case of ACRI occuring under
conditions relevant to VLF-emission in the magnetosphere, the energy
and pitch-angle changes of synchronous electrons may be enchanced by a
factor $10^2 \div 10^3$ in comparison with ones for nonsynchronous resonant
electrons. So the small in density group of synchronous particles may
give significant contribution to a whistler-mode wave damping in TPL.}
\end{quote}
\vskip1.2cm

\begin{center}
{\small\bf 1. INTRODUCTION}
\end{center}
\nopagebreak
\bigskip

The cyclotron resonance of charged particles with whistler-mode waves (CRI)
is one among the basic mechanisms which govern a dynamics of these waves
in the magnetosphere and in the ionosphere. As an example, CRI is closely
related to one of the most fascinating phenomena -- so-called VLF-emissions
triggering
in the magnetosphere (see, for example, review papers of {\sc Molchanov},
1985; {\sc Omura} {\it et al.}, 1991; {\sc Helliwell}, 1993 and
{\sc Rycroft}, 1993 and papers of {\sc Dysthe}, 1971; {\sc Karpman}
{\it et al.}, 1974; {\sc Nunn}, 1974; {\sc Dowden} {\it et al.}, 1978;
{\sc Matsumoto}, 1979; {\sc Bell}, 1984; {\sc Nunn}, 1984). Moreover, CRI
may give also the main contribution to VLF-waves damping and growth
and caused by them the pitch-angle diffusion of magnetospheric plasma hot
population to loss cone results to an anomalous precipitation of energetic
electrons from the radiation belts into the upper atmosphere. From practical
point of view, registration of VLF-emissions in the magnetosphere and
in the ionosphere and charged particle precipitation induced can be
applied to various diagnostic problems, for instance, to global control of
the geliogeophysical enviroment, to the forecasting of radiowaves
propagation in the Earth-ionosphere waveguide, to the satellite
monitoring of natural crisis processes like typhoons and earthquakes
and so on.

Specific feature of CRI in the magnetosphere is the key role of
inhomogeneouty of the geomagnetic field $\overrightarrow{B_0}$ and
the plasma density $n_e$ which determine the typical space scale
of cyclotron resonance region, its location along the geomagnetic
field line, the intensity and direction of energy transfer at the
nonlinear stage of a wave-particle interaction (WPI). The estimates
performed (e.g. {\sc Bell} and {\sc Inan}, 1981) have shown that
under the typical magnetospheric conditions the energetic electron
scattering by VLF-wave with significant changes of particle energy
and pitch-angle occurs only if the wave amplitude becomes large
enough because the typical space scale of CRI-region is usually
much less than the inhomogeneity length of geomagnetic field
$\overrightarrow{B_0}(S/L_B)$. Indeed, if $\omega$ and $k$ are
the whistler-mode wave frequency and wave vector respectively,
$\; Y=\omega_{He}/\omega\;$ is the energetic electron dimensionless
gyrofrequency, $S$ -- is the arc length along the geomagnetic field
$B_0(S/L_B)$ and $L_B$ is the inhomogeneity length, then one can
determine the small parameter of problem considered by the following
formulae
$$
\delta = 2\pi ~\frac{(Y-1)}{(1+\alpha_{\perp}^2)} \frac{(2Y+1)}{kL_B}
\  \sim \ 10^{-4}\div 10^{-5}
$$
where $\alpha_{\perp}^2 =v_{\perp}^2(Y-1)/[(2Y+1)\ v_R^2]$ -- is the
square of electron perpendicular velocity normalized on the typical
value, $v_R=(\omega-\omega_{He})/k$ is the resonance velocity.
According to the linear theory of WPI, the typical space scale of
cyclotron resonance region located outside the equatorial plane
is of the order of $\: {\it l}_R= L_B\delta^{1/2}$. At the equatorial
plane, the geomagnetic field gradient becomes zero. So in the case of WPI
located at the equatorial plane, the space scale of cyclotron resonance
region increases up to $\: {\it l}_R\equiv L_B\delta^{1/3}$. Nevertheless,
in both cases under conditions, typical for VLF-emissions generation
in the magnetosphere, the CRI-region space scale $\; {\it l}_R\,$ is
about two order of magnitude less than the inhomogeneity length $L_B$.
Therefore, it is of considerable importance to study the possibility
of sharp growth of CRI temporal duration, for instance, due to extent
of the interaction region space scale over the substantial portion of
the inhomogeneity length $L_B$.
For the cyclotron resonance interaction of energetic electrons
with the ducted whistler-mode wave of variable frequency in the equatorial
plane vicinity, this problem was considered by {\sc Brinca} (1981) and
{\sc Bell} and {\sc Inan} (1981). It was shown that the wave frequency
variation in the case of optimum frequency function allows to enchance
significantly the interaction region space scale for the most stable
trapped electrons. In the case of fixed-frequency wave this problem
was studied by {\sc Erokhin} (1995) and it was founded that the ACRI
takes place in the stationary transient boundary layer under consistency
of the whistler-mode wave amplitude $B_W$ with the magnetic field
gradient. So the following condition must be fulfilled : $\; B_W/B_0\sim
1/kYL_B\;$. This condition is in analogy with the long-lasting resonance
condition considered by {\sc Helliwell} (1967) and the second-order
resonance one described by {\sc Nunn} (1971) but in contrast to these
articles, paper of {\sc Erokhin} (1995) relates to so-called
synchronous particles whose phase $\Phi$ (its definition see below)
is close to $\pi /2$ during their crossing of TPL. As a result in the case
of synchronous particles the cyclotron resonance interaction becomes
a large-scale phenomena because the interaction region space scale
is comparable with the magnetic field inhomogeneity length. In
TPL it is observed the antidrift dynamics of synchronous particles
and relative changes of their energy and pitch-angles are of the order of
$100$ percents if the magnetic field variation is large enough $\,
\delta B_0\sim B_0\, $. Consequently, in the transient plasma layer
the anomalous cyclotron resonance interaction of synchronous
particles with the whistler-mode wave takes place and one would
expect ACRI to modify the wave damping.

As both rising and falling tones are observed in the magnetosphere
when triggering VLF-emission, it is necessary to study the wave
frequency variation influence on the anomalous cyclotron resonance
interaction of energetic electrons with the ducted whistler-mode
wave in the transient plasma layer. The present paper is devoted to
solving this problem. Its solution allows to perform correct estimates
of the wave damping for anomalous CRI in TPL.

The paper structure is the following. The basic equations derivation
and relations resulted are given in Section 2. The case of fixed
wave frequency is described in Section 3. Analytical and numerical
results of studying the frequency sweeping influence on anomalous
CRI in the stationary TPL are given in Section 4. Results obtained
are discussed in Section 5.
\vskip1.2cm

\begin{center}
{\small\bf 2. BASIC EQUATIONS.}
\end{center}
\nopagebreak
\bigskip

Let us consider the cyclotron resonance interaction between energetic
electrons and the ducted wistler with a frequency $\; \omega<\omega_{Be}
\;$ propagating along a weakly inhomogeneous both plasma and
magnetic field $\overrightarrow{B_0}$. As this interaction is
localized in the vicinity of field line it is quite natural to use the
curvilinear orthogonal coordinate system with the basic vector along the
arc length $S$ of the field line, normal and binormal to this line. It is
convenient to introduce the dimentionless variables $s=\omega S/c,\quad
t'=\omega t,\quad \vec{\beta}=\vec{v}/c,\; $ where $\vec{v}$ - is the
velocity of electron.

The equations of motion for the non-relativistic electrons
mirroring in the magnetic field $\overrightarrow{B_0}$ took
the standard form (see, for example, Dysthe,1971; Nunn, 1974):

$$
{\displaystyle\frac{d\beta_{\|}}{dt'}}=\Omega_W \beta_{\perp}\sin{\Phi}-
\beta_{\perp}^2 {\displaystyle \frac{Y_s}{2Y}} ,
\qquad {\displaystyle\frac{ds}{dt'}}=\beta_{\|}.
\eqno(1)
$$

$$
{\displaystyle\frac{d\beta_{\perp}}{dt'}}=\Omega_W (\beta_{ph}-\beta_{\|})
\sin{\Phi} +\beta_{\|}\beta_{\perp}{\displaystyle \frac{Y_s}{2Y}} ,
$$

$$
{\displaystyle\frac{d\Phi}{dt'}}={\displaystyle\frac{\beta_{\|}-
\beta_R}{\beta_{ph}}}+{\displaystyle\frac{\Omega_W}{\beta_{\perp}}}
(\beta_{ph}-\beta_{\|})\cos{\Phi}-(1-{\displaystyle\frac{\beta_{\|}}
{\beta_{g}}}){\displaystyle\frac{\delta \omega}{\omega}},
$$

Here $\;\Omega_W=|e|B_W/m_ec\omega\;$ -- is the dimensionless wistler
amplitude, $\;\beta_{ph}=(\omega /\omega_{pe})(Y-1)^{1/2}\,$ and
$\,\beta_g =-2\beta_R /Y\,$ -- phase and group whistler mode wave
velocities respectively, $\,\beta_R=(1-Y)\beta_{ph}\,$ --resonance
velocity, $Y_s\equiv\partial_sY$ -- magnetic field gradient, $\,\Phi$ --
the complement of the angle between the electron\'s perpendicular
velocity $\vec{v}_{\perp}$ and $\vec{B}_W$. The slow frequency sweeping
rate of the whistler is taken into account by term $\delta\omega/\omega$
in the wave phase equation, and under wave amplitude diffusion
neglecting it depends only on

$$
t_g'\equiv t'-\int\limits_0^s \frac{ds'}{\beta_g(s')}.
$$

Having in mind the VLF-emissions magnetospheric typical parameters
one can assume the whistler-mode wave frequency change to be small
enough ({\it i.e.} $|\delta\omega |\ll \omega$) during the particle 
crossing the resonance region.

It is nesesary to pay attention to the following circumstance. As far
as we are interesting in the WPI large-scale dynamics when the resonance
region is global in a size comparable with the inhomogeneity length
of static magnetic field $B_0$, in the equation $(1)$ for phase $\Phi$
we should keep the second term proportional to the whistler
amplitude $\Omega_W$. In the case of small-scale WPI analysis,
it is usually neglected (see, for example, {\sc Matsumoto}, 1979;
{\sc Brinca}, 1981).

According to {\sc Erokhin} (1995) and {\sc Erokhin} {\it et al.}
(1995), the magnetic field and the plasma density in the transient
plasma layer are monotonous functions of the arc length $S$. So
it is possible to put them in one to one correspondence $\, \omega_{pe}
=\omega_{pe}(Y)\,$, where $\omega_{pe}$ is the electron langmuir frequency.
To concretize the following calculations we are using the power
function of the type $\; \omega_{pe}(Y)=\omega_{pe}(Y_0)(Y/Y_0)^\sigma\;$,
where $Y_0$ is the magnetic field at some point $S_0$ inside TPL.
For the power index $\sigma$ it is usually taken the value $\sigma=0$
(DE-model) or $\sigma=0.5$ (CL-model; see, for example, {\sc Bell}
and {\sc Inan}, 1981). The spatial dependence of whistler amplitude 
$\Omega_W$ can be given by the condition of energy flux conservation in
the ray tube which crossection is inversely with $B_0$ (see, for
example, {\sc Molchanov}, 1985). Therefore we obtain the following
scaling of the cyclotron resonance and phase velocities as well as
the whistler amplitude on the magnetic field $Y$ :

$$
\beta_{ph}=\frac{\beta_*}{Q^2(Y)},~~~~~~\beta_R(Y)=-\beta_*
\frac{(Y-1)}{Q^2(Y)},
\eqno(2)
$$
$$
\Omega_W(Y)=\Omega_*Y^{1/2}Q(Y),~~~~~~Q(Y)\equiv \frac{Y^{\sigma /2}}
{(Y-1)^{1/4}},
$$
where the following notations were used
$$
\beta_*\equiv {\beta_{ph}(Y_0)}{Q^2(Y_0)},~~~~~~\Omega_*\equiv
\frac{\Omega_W(Y_0)}{Y_0^{1/2}Q(Y_0)}.
$$

For convenience of the subsequent analysis it is nesessary to
transform the set of equations (1) to canonical form. Let us introduce
the typical space scale of TPL as $L_t$. In dimensionless variable
$s$ it is equal to $\; s_*\equiv \omega L_t/c =\beta_*/2\Omega_*\gg 1\;$.
Applying the scale transformations of variables $\; \beta_{\|}=\beta_*u,
\;\; \beta_{\perp}=\beta_*v,\;\; s=s_*\xi,\;\; t'=\tau/2\Omega_* \;$ 
to the set of equations (1) and taking into account (2) we reduce (1) 
to the canonical form:

$$
2{\displaystyle\frac{du}{d\tau}}=Q(Y)Y^{1/2}v\sin{\Phi}-v^2F(Y),
\qquad{\displaystyle\frac{d\xi}{d\tau}}=u\; ,
$$
$$
2{\displaystyle\frac{dv}{d\tau}}={\displaystyle\frac{Y^{1/2}}{Q(Y)}}
[1-uQ^2(Y)]\sin{\Phi} +uvF(Y),
\eqno(3)
$$
$$
2{\displaystyle\frac{d\Phi}{d\tau}}=\chi [uQ^2(Y)+Y-1]+
{\displaystyle\frac{Y^{1/2}}{Q(Y)v}}[1-uQ^2(Y)]\cos{\Phi} -
 \chi (1-{\displaystyle\frac{u}{u_g}})
{\displaystyle\frac{\delta \omega}{\omega}},
$$
where $\quad \chi =(1/\Omega_*)\gg 1,\quad u_g=2(Y-1)/Q^2(Y)Y\;$ and
$\; F(Y)\equiv Y_{\xi}/Y\;$ is the magnetic field logarithmic gradient.
Eqs. (3) form the basic set of equations for the subsequent analysis 
and the function $Y(\xi)$ determines the TPL spatial structure.
It will be founded below from the condition of the synchronous
particles existence. To describe briefly the TPL spatial structure
in the case of constant wave frequency one puts $\delta\omega =0$ in
Eqs. (3). According to the paper by {\sc Erokhin} {\it et~al.} (1995),
the anomalous cyclotron resonance interaction of energetic electrons
with the ducted whistler-mode wave takes place for the group of 
synchronous particles defined by the following conditions: 1) the
phase $\Phi$ along the synchronous particle path is constant and is
equal to $\;\Phi_s=\pi /2$;$\;$ 2) the parallel velocity of synchronous
particle $u_s$ is equal to the cyclotron resonance velocity $\; u_R
\equiv -(Y-1)/Q^2(Y)\;$. Putting in (3) $\;u=u_R,\; \Phi =\pi /2\;$
and $\delta\omega =0$ we obtain the first order nonlinear equation
for the magnetic field profile in TPL:
$$
\frac{dY}{d\xi}\equiv YF(Y)=\frac{Y^{\mu}(Y-1)^{1/4}[\Lambda -R(Y)]^{1/2}}
{\Lambda +2(1-\sigma +\sigma /Y)R(Y)}
\eqno(4)
$$
where $\quad \mu =1+\sigma /2,\quad R(Y)\equiv u_R^2(Y)$ and $\Lambda$
is a positive parameter determining the one-parameter set of the magnetic 
field profiles $Y(\Lambda,\xi)$, for which the anomalous CRI of energetic 
particles with the small amplitude whistler-mode wave may occur at the
entire transient plasma layer.

Let the power index $\sigma$ be in the range $\; 0\le\sigma\le 3/2\;$.
The solution of equation (4) exists if and only if the function
$Y(\Lambda,\xi)$ is in the region $\, (1,Y_m(\Lambda ))\,$, where 
$Y_m(\Lambda)$ is the single root of equation $\; \Lambda=R(Y_m)\;$,
monotonously increasing under parameter $\Lambda$ growth.

To determine $Y(\xi)$ we suppose the magnetic field in TPL to be
varying in the range $\; Y_1\le Y(\xi )\le Y_2\;$, where $\, Y_1>1\,$
 and $\, Y_2<Y_m(\Lambda )\,$. Then the magnetic field profile $Y(\xi )$
can be obtained by inversion of the following monotonous function
$$
\xi (Y)=\int\limits_{Y_1}^Y \; {\displaystyle\frac{[\Lambda +2(1-\sigma +
{\displaystyle\frac{\sigma}{x}})R(x)]\; dx}{x^{\mu}(x-1)^{1/4}
[\Lambda -R(x)]^{1/2}}}.
\eqno(5)
$$
Fig.1a depicts the part of profile (5) in the case when the magnetic
field is varying in the range $\; 2\le Y(\xi )\le 4\;$, the power
index corresponds to DE-model ($\sigma =0.5$) and the parameter 
$\Lambda$ takes the following values: $6.755$, $18$ and $40$. 
For the given values of parameters $\sigma$ and $\Lambda$ 
possible maximum magnitudes of the magnetic field $Y_m$ are
respectively: $4.001$, $5.674$ and $7.755$.

Fig.1b depicts the profile (5) with magnetic field variation in the
range $\; 2\le Y(\xi )\le 4\;$ with $\Lambda =40$ and different
values of power index $\sigma$. According to Fig.1b for the given
value of parameter $\Lambda$, the TPL width is larger in the case of 
DE-model.

For the transient plasma layer with magnetic field variation in the
range $\; Y_1\le Y(\xi )\le Y_2\;$, the layer width $l_{\xi}$ is
determined by the following expression:
$$
l_{\xi}=\int\limits_{Y_1}^{Y_2}\; {\displaystyle\frac{[\Lambda +2(1-\sigma +
{\displaystyle\frac{\sigma}{Y}})R(Y)]\; dY}{Y^{\mu}(Y-1)^{1/4}
[\Lambda -R(Y)]^{1/2}}}.
\eqno(6)
$$

\begin{center}
\vspace{0.5cm}

\mbox{\epsfysize=2in \epsffile{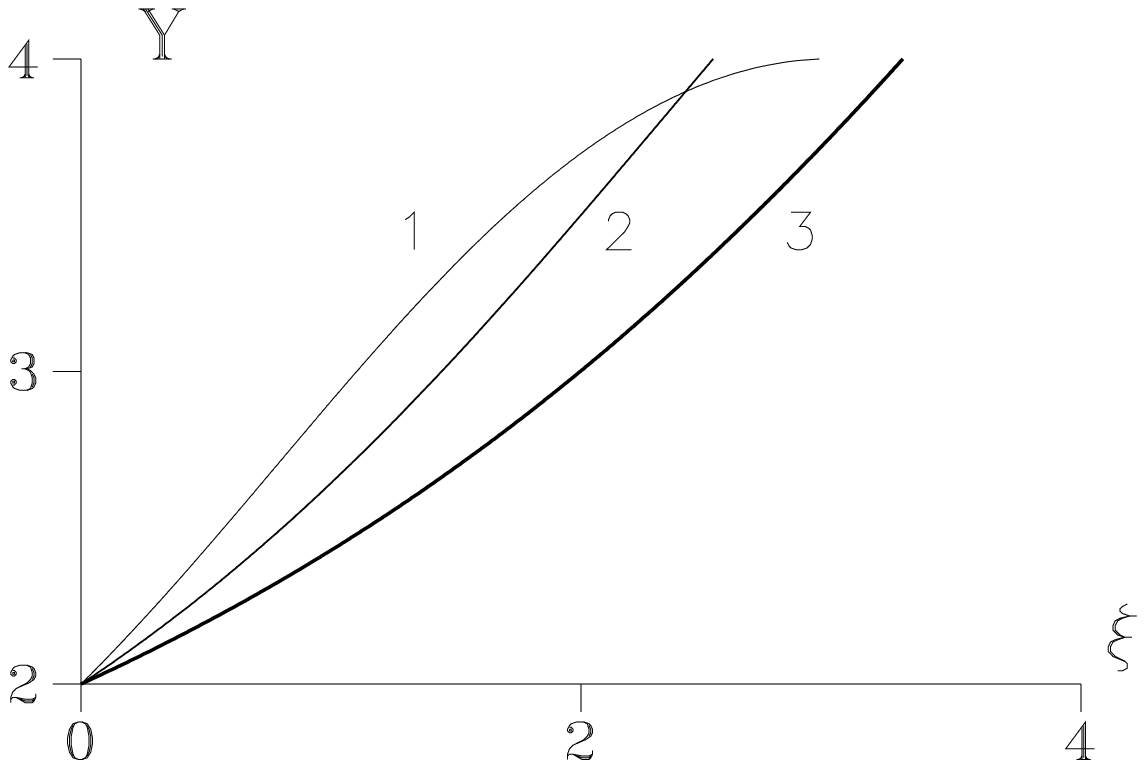}}

\end{center}

\begin{quote}
{\small Fig.1a. Magnetic field profile $Y(\xi )$ in the transient plasma layer for
$\sigma =0.5$ and different values of parameter $\Lambda$:
1 -- $\;\Lambda =6.755$; 2 -- $\;\Lambda =18$; 3 -- $\;\Lambda =40$.
TPL corresponds to the dispersed magnetic field jump $\, 2\le Y(\xi )
\le 4\,$.}  
\end{quote}
\begin{center}
\vspace{0.5cm}

\mbox{\epsfysize=2in \epsffile{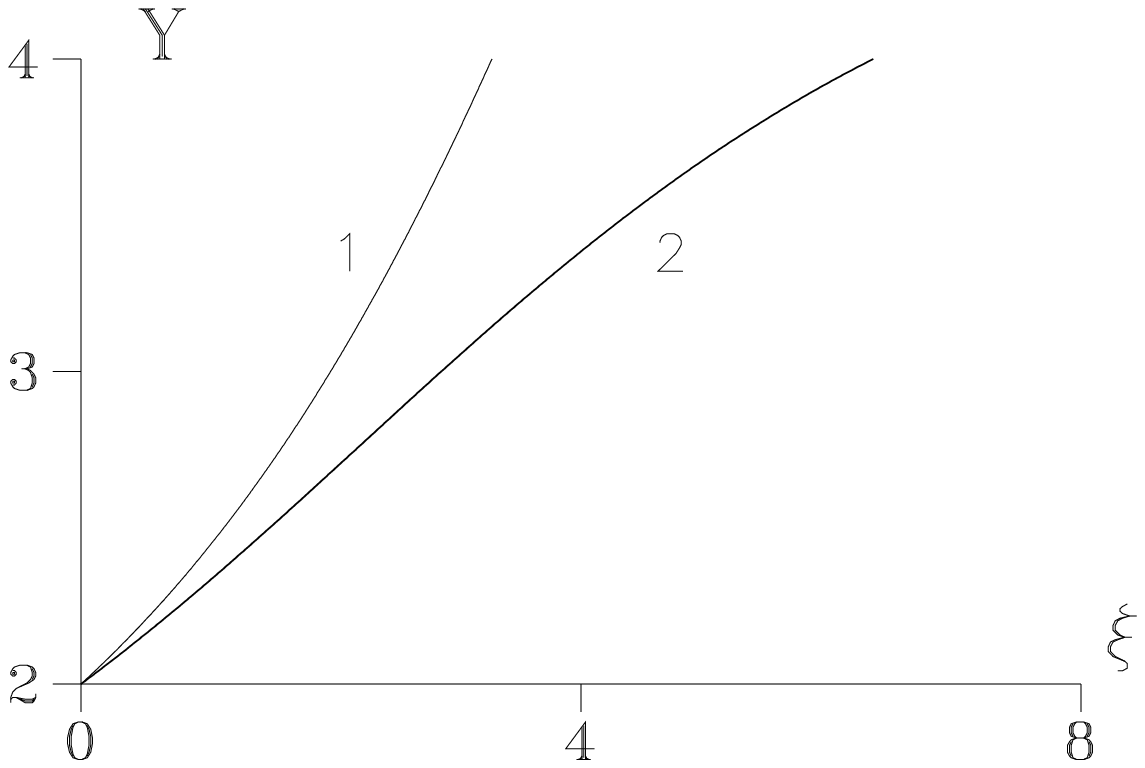}}

\end{center}
\begin{quote}
{\small Fig.1b. Magnetic field profile in the case $\, 2\le Y(\xi )
\le 4\,$ for $\;\Lambda =40\;$ and different values of power index
$\sigma$: 1 -- $\,\sigma =0.5$ (CL-model); 2 -- $\,\sigma =0$
(DE-model).}  
\end{quote}
\bigskip
or in the dimensional variable $S$ we obtain $\; L_S=(c\beta_*/2
\omega\Omega_*)\, l_{\xi}\;$. Fig.1c illustrates the dependence of
TPL-width $l_{\xi}$ on parameter $\Lambda$, defined by formula (6)
in the case of $\; Y_1=1,\;\; Y_2=Y_m(\Lambda)\;$ for two values of the
power index $\sigma$. It can be seen again that the layer width is 
larger for DE-model of the plasma density. As far as for synchronous 
particles there is a well known integral of motion ({\sc Karpman}
{\it et al.}, 1974; {\sc Nunn}, 1974)
$$
u_s^2+(1-\frac{1}{Y})v_s^2=\Lambda\; ,
$$
their energy ${\cal E}_s$, magnetic moment $\mu_s$ and velocities components 
determined by the magnetic field local strength
$$
{\cal E}_s(Y)=\frac{Y\Lambda -R(Y)}{2\, (Y\! -1)},~~~~~\mu_s(Y)\equiv
\frac{v_s^2(Y)}{2\, Y}=\frac{\Lambda -R(Y)}{2\, (Y\! -1)},
\eqno(7)
$$
$$
u_s(Y)=-\frac{(Y-1)^{3/2}}{Y^{\sigma}},~~~~~v_s(Y)=\left[\frac{Y}{(Y-1)}
(\Lambda -R(Y))\right]^{1/2}.
$$
\begin{center}
\vspace{0.5cm}

\mbox{\epsfysize=2.5in \epsffile{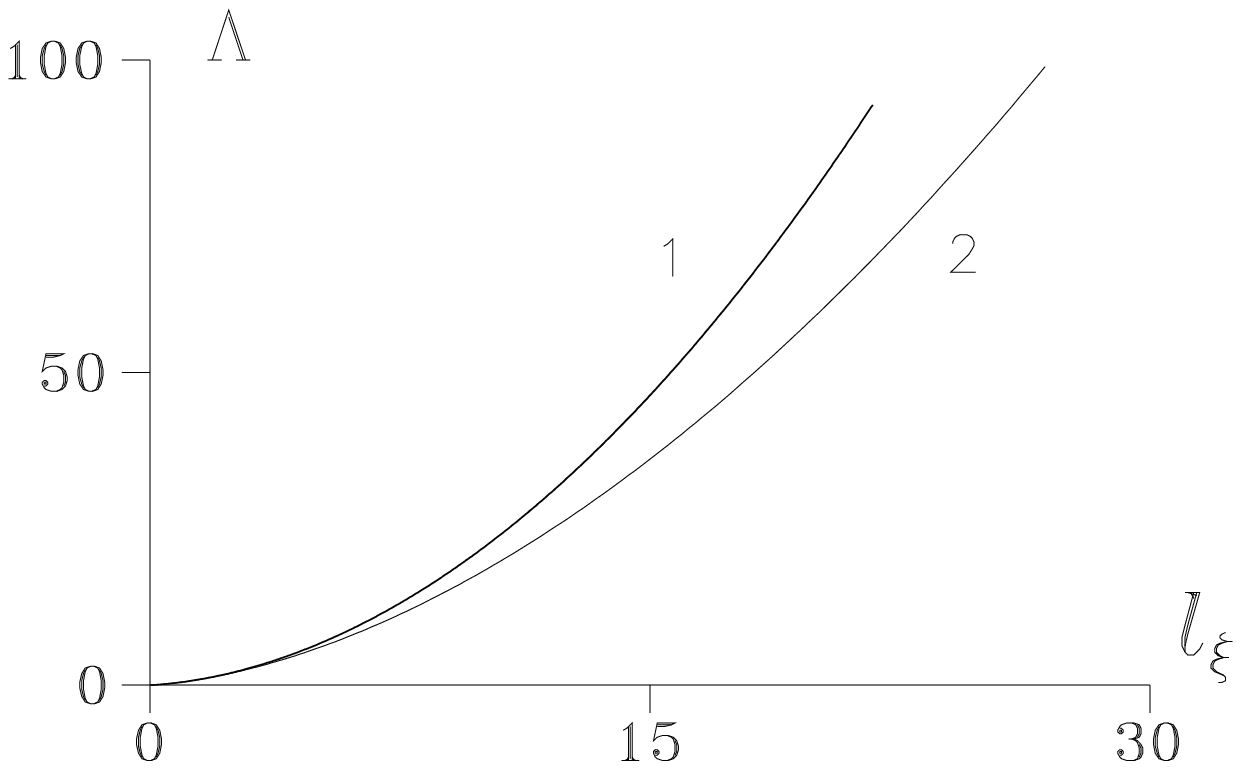}}

\end{center}
\begin{quote}
{\small Fig.1c. Dependence of TPL-thickness $\, {\it l}_{\,\xi}(\Lambda )\,$
with magnetic field variation $\,1\le Y(\xi )\le Y_m(\Lambda )\,$ on
parameter $\,\Lambda\,$ in the cases: 1 -- $\;\sigma =0.5$; 2 -- }
$\;\sigma =0$
\end{quote}
\bigskip

The time $\Delta\tau_s$, required for TPL-crossing by the synchronous
particle, is equal to
$$
\Delta\tau_s=\int\limits_{Y_1}^{Y_2} \frac{dY}{YF(Y)\; u_g(Y)}\; .
$$

Let us consider again equations (3). To perform the subsequent
calculations we assume the following model of wave frequency
variation $\; \delta\omega /\omega =h(p)\;$, where h(p) is
the given function. The spatio-temporal dependence of function
$p(\xi,\tau )$ is governed by the following equation $\; dp/d\tau =
\nu\, (1-u/u_g)\;$, where parameter $\nu$ determines the 
frequency sweeping rate. The frequency sweeping rate can be
characterized by a function $\; \Gamma\equiv (\partial f/\partial t)/f^2
\;$ which has the folowing scaling
$$
\Gamma\simeq 4\cdot 10^{-5}\left[\frac{\partial f/\partial t}
{1\mbox{ kHz/sec}}\right]\,\left[\frac{5\mbox{ kHz}}{f}\right]^2\, ,
$$
orientated towards the magnetospheric VLF-emissions parameters.
Functions $\Gamma$ and $h$ are related by $\; dh/dp=\chi\Gamma/4\pi\nu \;$.
Therefore in the case of constant wave frequency sweeping rate,
corresponding to the choice $h(p)=p$, parameter $\nu$ is equal to
$\; \nu =\chi\Gamma/4\pi\;$. So under the typical conditions of 
VLF-emissions in the magnetosphere, one has $\nu <1$.
\vskip1.2cm

\begin{center}
{\small\bf 3. ESSENTIAL FEATURES OF THE ANOMALOUS CYCLOTRON
RESONANCE INTERACTION OF ENERGETIC ELECTRONS WITH THE FIXED
FREQUENCY WHISTLER-MODE WAVE IN TPL.}
\end{center}
\nopagebreak
\bigskip

Before the studying the whistler frequency drift influence on the
anomalous CRI in TPL, to clarify the following analysis it is nesessary 
to describe shortly the case of fixed frequency wave coresponding
to the condition $\;\delta\omega =0\;$ in Eqs. (3). Let us assume
that in stationary TPL the magnetic field varies in the range
$\; Y_1\le Y(\xi )\le Y_2\;$ and the whistler-mode wave propagates
in the direction of $Y(\xi )$ growth. Therefore, the resonant 
particles are travelling in the opposite direction towards the
wave. The maximum cyclotron resonance interaction in the transient 
plasma layer takes place for the group of synchronous electrons
whose phases $\Phi_s$ are constant and equal $\pi /2$ and other
parameters are defined by (7). The group of synchronous particles
can also be defined in the different way. Let us introduce
function $J$ by the following expression:
$$
J=u^2+(1-\frac{1}{Y})\,v^2\; .
$$

According to papers of {\sc Karpman} {\it et al.} (1974) and {\sc Nunn}
 (1974), for resonant particles function $J$ is the approximate
integral of motion. Consequently, the group of synchronous particles
can be defined by the folowing conditions: $\; u_s\approx u_R,\quad 
\Phi_s\approx \pi /2\;$ and $\; J\approx \Lambda \;$. Nonconserving
part of the approximate integral of motion $J$ can be easily
estimated from (3). Performing the asymptotic integration, we
obtain the modified integral of motion
$$
I=J+\frac{2v\,Q\,\chi}{\sqrt Y}\,\cos{\Phi}\; .
$$

In turn, non conservation of the function $I$ is caused by fastly
oscillating terms of the order of $\, 1/\chi^2\,$ but the particle 
crossing of the local hyroresonance region is accompanied by the
jump of function $I$ proportional to $\, 1/\chi^{3/2}\,$.

Let us turn back to formulae (7). According to (7) when crossing 
the TPL the synchronous particles increase their energy, magnetic
moment, perpendicular velocity and pitch-angle $\; \alpha_s =
\arctan{(v_s/|u_s|)}\equiv \tan^{-1}{(v_s/|u_s|)}\;$. At the same time,
when moving in the direction of magnetic field decreasing, their
parallel velocity goes down, {\it i.e.} in the transient plasma
layer the antidrift dynamics of synchronous particles takes place.
Earlier similar effects for electrons, trapped by the whistler-mode
wave in the equatorial plane vicinity, were pointed out by {\sc
Matsumoto} (1979).

Now let us assume that at TPL-entrance the velocities of incoming
electrons coincides with the synchronous particle velocity, {\it i.e.}
$u(0) = u_s(Y_2)\;$, $v(0) = v_s(Y_2)\;$, but there is a small deviation 
$\theta_0$ in the initial phase $\;\Phi (0)=\Phi_s+\theta_0\;$.

Numerical calculations and analytical estimates show us that there are
some constants $\,c_{(-)}<0<c_{(+)}\,$ so that for the initial
phase detuning $\,c_{(-)}/\chi <\theta_0<c_{(+)}/\chi\,$ during
the resonance particle pass TPL, its phase is confined in the range
$\, 0<\Phi <\pi\,$ and its energy and pitch angle changes are
close to the synchronous particle ones. If the initial phase detuning
is large $\;\theta_0 >c_{(+)}/\chi\,$ or $\,\theta_0 <c_{(-)}/\chi\;$,
the duration of resonance interaction $\Delta\tau_R$ becomes less
than the time required for the resonance particle to cross the 
transient plasma layer. Under the growth of $\theta_0$ the time
$\Delta\tau_R$ decreases as $\;\Delta\tau_R\sim 1/(\chi\,|\theta_0|)^
{1/2}\;$. So there is the following scaling of the resonance particle 
energy growth $\Delta{\cal E}$ in dependence on the initial phase
detuning $\theta_0\,$: $\;\Delta{\cal E}\sim 1/(\chi\,|\theta_0|)^{1/2}\;$.
Fig.2a depicts the chart of the resonance particle relative energy
growth $\; A\equiv 10^2\Delta{\cal E}/{\cal E}_0\;$ in dependence
on the normalized initial phase detuning $\chi\theta_0$ for small
$\theta_0$. The system (3) was integrated numerically for TPL with
the magnetic field variation in the range $\;2\le Y(\xi )\le 4\;$
and the following parameters: $\,\sigma =0.5,\; \Lambda =10,\; \nu=0\,$.
The initial data correspond to the synchronous particle {\it i.e.}
$u_0=u_s(Y_2),\; v_0=v_s(Y_2)\,$, where $\, Y_2=4\,$. According to
Fig.2a the maximum cyclotron resonance interaction of energetic 
electrons with the whistler mode of constant wave fre- 
\begin{center}
\vspace{0.5cm}

\mbox{\epsfysize=2.5in \epsffile{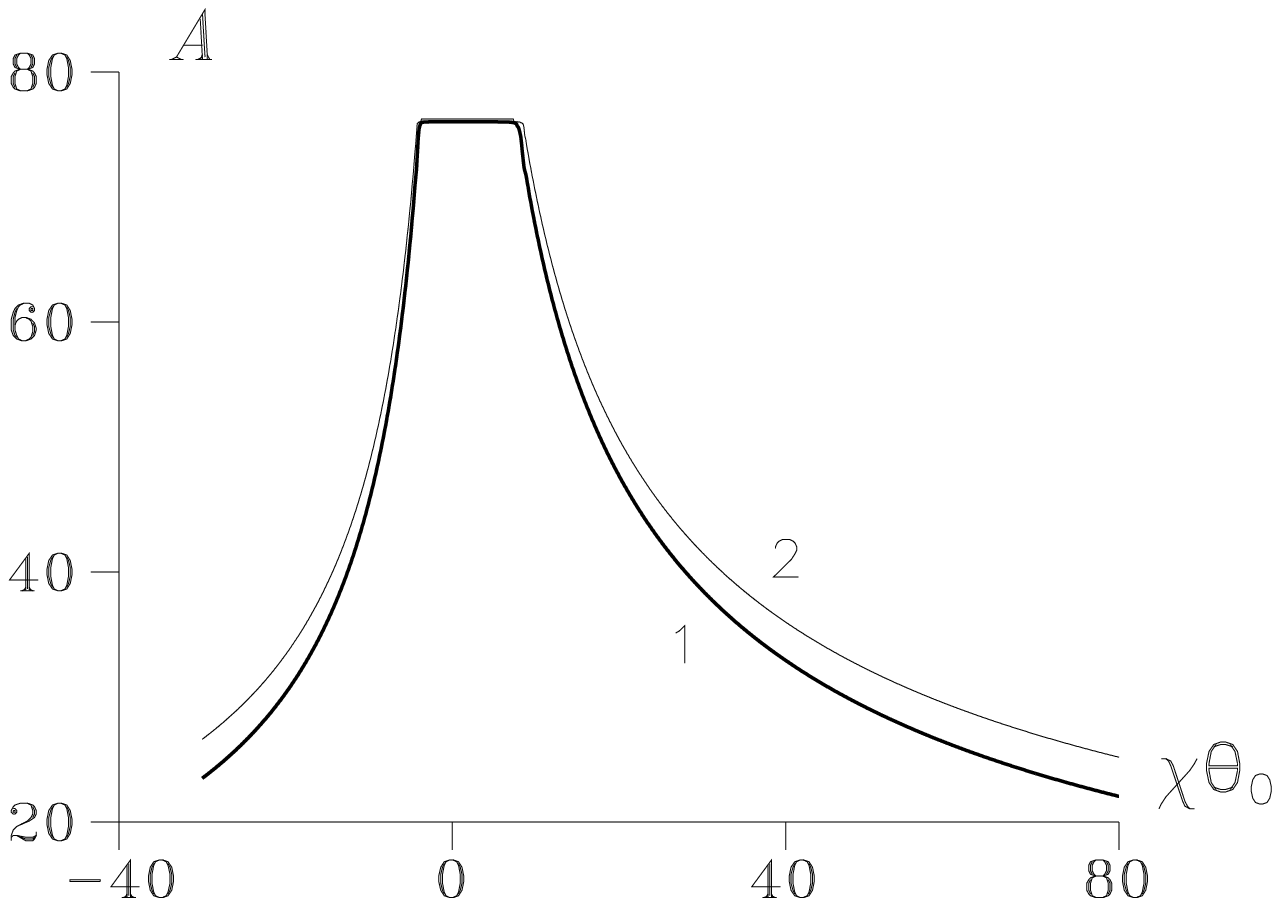}}

\end{center}
\begin{quote}
{\small Fig.2a. Dependence of synchronous particle energy gain $A$
on normalized initial phase detuning $\,\chi\theta_0\,$: 1 --
$\;\chi =10^3$; 2 -- $\;\chi =10^4$.}
\end{quote}
\bigskip
quency takes 
place for the small initial phase detuning $\; |\theta_0|\sim 1/\chi\;$.
It can be seen that charts are nonsymmetric on the phase detuning
$\theta_0$ and the resonant particle is gaining more energy in the
case of positive $\theta_0$. In range $\;\chi |\theta_0|\ge 10^2\;$
the dependence $A$ on the parameter $\chi$ becomes appreciable.
Similar effect is also observed for the pitch angle scattering
of synchronous particles in TPL. Let $\Delta\alpha$ represent the
synchronous particle pitch angle change during the cyclotron resonznce
interaction with wistler in TPL. We characterize the efficiency of
electron pitch angle scattering during resonant WPI by function
$\; B=10^2(\Delta\alpha /\alpha_0)\;$, where $\alpha_0$ is the
electron pitch angle at the resonant region entrance. Notice that
in theory of the VLF-emissions triggering in the magnetosphere the
pitch angle scattering of energetic electrons by whistler-mode wave
is usually characterized by the change of equatorial pitch angle
$\;\alpha_0 =\sin^{-1}[(Y_e\,\sin^2\!\alpha\, /\, Y)^{1/2}]\;$, where
$Y_e$ is the magnetic field strength at the equatorial plane. The
dependence of $B$ on the normalized initial phase detuning $\chi
\theta_0$ is shown in Fig.2b for problem parameters as in Fig.2a.
\begin{center}
\vspace{0.5cm}

\mbox{\epsfysize=2.5in \epsffile{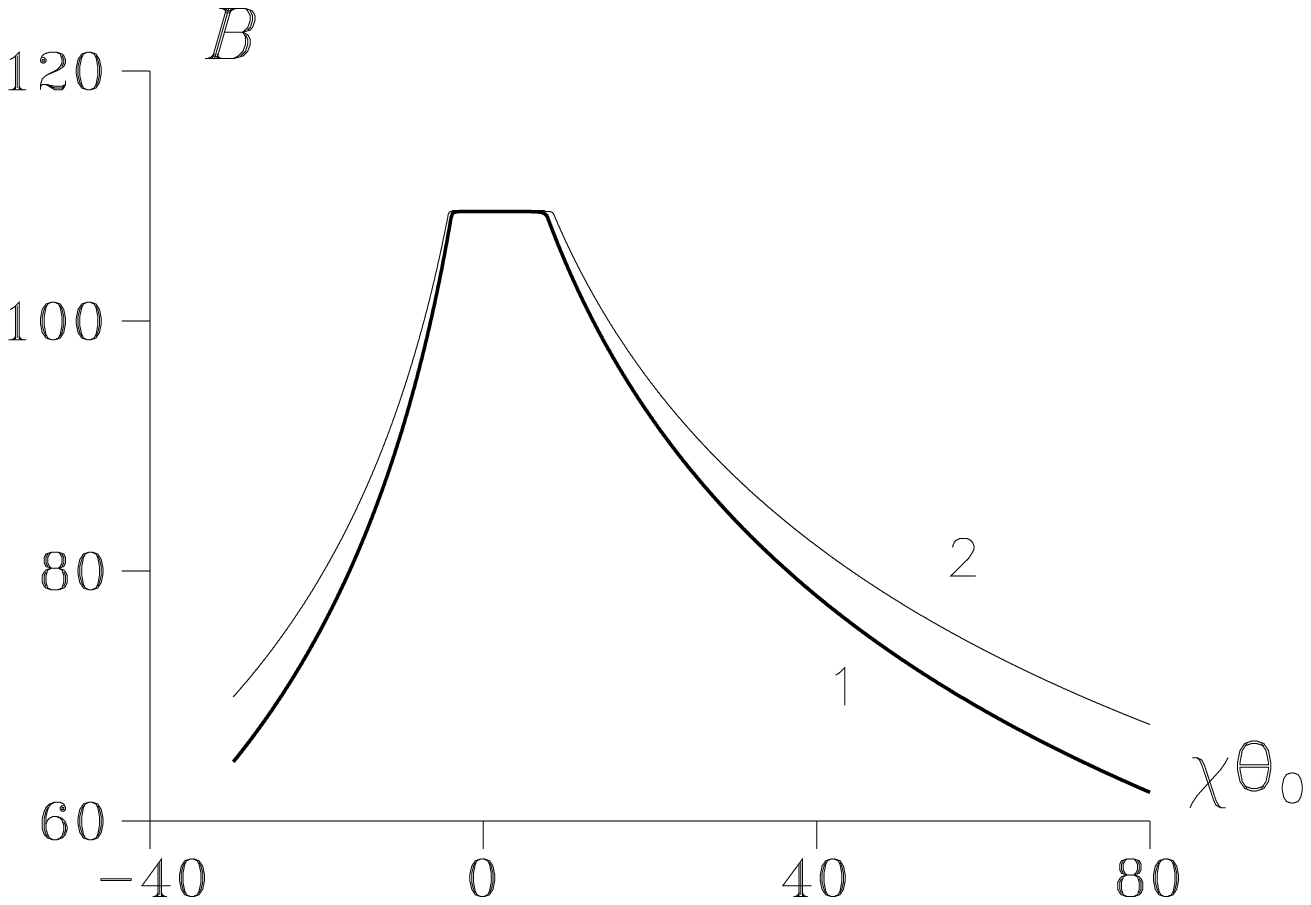}}

\end{center}
\begin{quote}
{\small Fig.2b. Dependence of pitch angle scattering efficiency $B$ of 
synchronous particle on normalized initial phase detuning $\,\chi\theta_0
\,$: 1 -- $\;\chi =10^3$; 2 -- $\;\chi =10^4$.}
\end{quote}
It can be seen that the pitch angle scattering efficiency $B$
is larger approximately by factor $1.5$ of the energetic efficiency
$A$. From charts of Fig.2a and Fig.2b it follows that the cyclotron
resonance interaction of synchronous particles, characterized by
small $\theta_0$, with the wistler-mode wave in TPL is anomalously
strong because of their energies and pitch angles relative changes
are about $100\%$. Under $\theta_0$ growth the efficiency of cyclotron 
resonance interaction falls. In the range of $\, |\theta_0|\sim 1\,$
the resonance region space scale is small enough. So with accuracy
of a numerical coefficient of the order of unity one can use the
linear estimate of $A$
$$
|A|\le\frac{\sin{2\alpha_0}}{\sqrt{\chi}}\frac{Q^{3/2}}{(\Lambda-Y)^{1/4}}
\left[\frac{\Lambda +2\,(1-\sigma+\sigma/Y)\, R}
{3-2\sigma +2\, \sigma/Y}\right]^{1/2}
\frac{1}{(Y-1)^{9/4}}
\eqno(8)
$$

Taking $\,\chi=10^4\,$ and other parameters values correspondingly
to Fig.2 we obtain from (8) that $\, |A|\le 0.2\,$ {\it i.e.} for
large $\theta_0$ the relative change of resonance particle energy is
about 400 times less than the synchronous particle one.

Results of numerical calculations of function $A$ obtained by integrating 
system (3) in the case of large phase detunings $\theta_0$ are given
in Fig.2c. The problem parameters are the 
\begin{center}
\vspace{0.5cm}

\mbox{\epsfysize=2in \epsffile{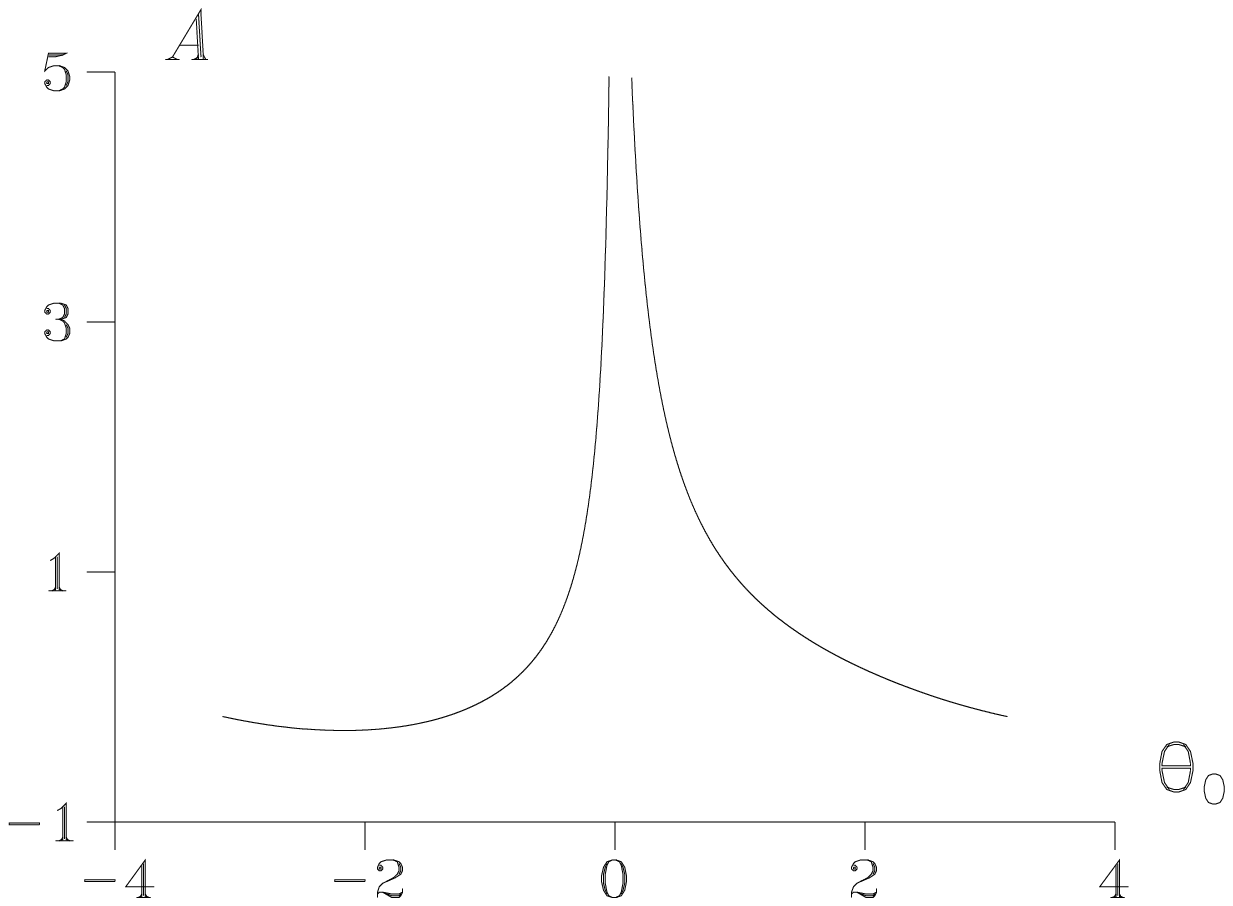}}

\end{center}
\begin{quote}
{\small Fig.2c. Dependence of resonance particle energy gain $A$
on initial phase $\theta_0$ in the case of large $\theta_0$. The problem
parameters are the following: $\;\chi =10^4,\;\nu =0,\;\Lambda =10,\;
\sigma =0.5,\; Y_0=4,\; u_0=u_s(Y_0)\,$ and $\; v_0=v_s(Y_0)\,$.}
\end{quote}
same as for Fig.2a and
$\,\chi=10^4\,$. Notice the narrow peak of the energetic efficiency
$A$ for small $\theta_0$ corresponding to the synchronous particles.
For comparison let us consider the cyclotron interaction of trapped
electrons with the wistler-mode wave in TPL. The trapped particles 
perpendicular velocities at the TPL-entrance should be different 
the synchronous particle one $v_s(Y_0)$. Introduce the normalized
perpendicular velocity $\,\ae\equiv v/v_s(Y)\,$ and functions $g(Y)$
and $\,\rho (Y,\ae )\,$:
$$
g(Y)=\frac{\Lambda-R(Y)}{\Lambda +2\,(1-\sigma+\sigma/Y)\, R}\quad ,
\eqno(9)
$$
$$
\rho =\ae \; g(Y)+\frac{1-g(Y)}{\ae}\quad ,
$$
where condition $\,0\le g\le 1\,$ takes place. From formula (9) it
follows that in dependence on the variable $\ae$ the function $\rho$
has minimum at $\;\ae =\ae_*(Y)\equiv\sqrt{(1/g)-1}\;$ which is equal
to $\; \min{\rho}\equiv\rho_* = 2\,\sqrt{g(1-g)}\le 1\;$. The trapped
particle dynamics corresponds to the motion of nonlinear oscillator
$\theta$ in the potential well $\; U(\theta )=\Omega_b^2\,(\rho\,\theta -
\sin{\theta})\;$, where $\;\Omega_b^2=\chi\,\ae\, v_s(Y)\,Y^{1/2}\,Q^3(Y)
\,/\,4\;$ is the bounce frequency square. The potential well exists
only for $\,\rho <1\,$. In the case of $\, 0<g<1/2\,$ the condition
$\,\rho <1\,$ is fulfilled for $\ae$ in the range $\, 1<\ae <\ae_*^2\,$.
If $g(Y)$ is ranged as $\,0.5<g<1\,$, the trapped particle perpendicular
velocity is less than the synchronous particle one, {\it i.e.} $\ae$
is in the interval $\,\ae_*^2<\ae <1\,$. The most long CRI of trapped
particles with whistler in TPL takes place for the initial phase
$\;\theta_0 =\arccos \rho (Y_0)\equiv\theta_b\;$ and $\,\ae =\ae_*\,$
when the trapped particle region has a maximum size in the phase
plane $\, (\theta ,u)\,$.

The numerical calculations of the cyclotron resonance interaction of 
whistler-mode wave with the stably trapped particles with energy levels
located closely to the potential well bottom were made, and the 
dependence of energetic efficiency $A$ on the trapped particles
perpendicular velocity $\,\ae_0\equiv\ae (Y_0)\,$ was studied. 
Fig.3a depicts the chart of 
\begin{center}
\vspace{0.5cm}

\mbox{\epsfysize=2in \epsffile{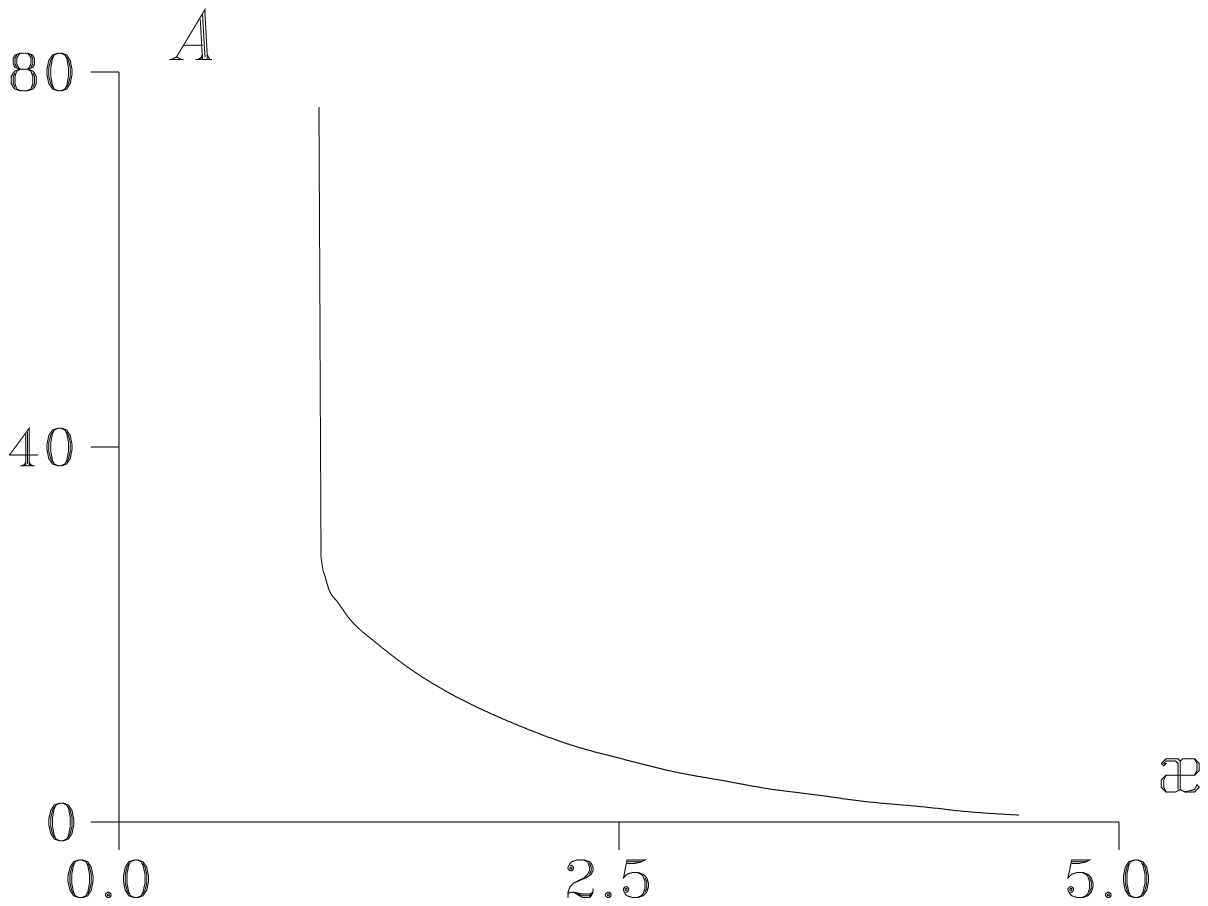}}

\end{center}
\begin{quote}
{\small Fig.3a. Dependence of energy gain of trapped particle, located
closely to potential well bottom, on normalized perpendicular velocity
$\,\ae_0$. Initial phase is equal to $\;\theta_0 =\cos^{-1}\rho (\ae_0 )$.}
\end{quote}
\begin{center}
\vspace{0.5cm}

\mbox{\epsfysize=2.5in \epsffile{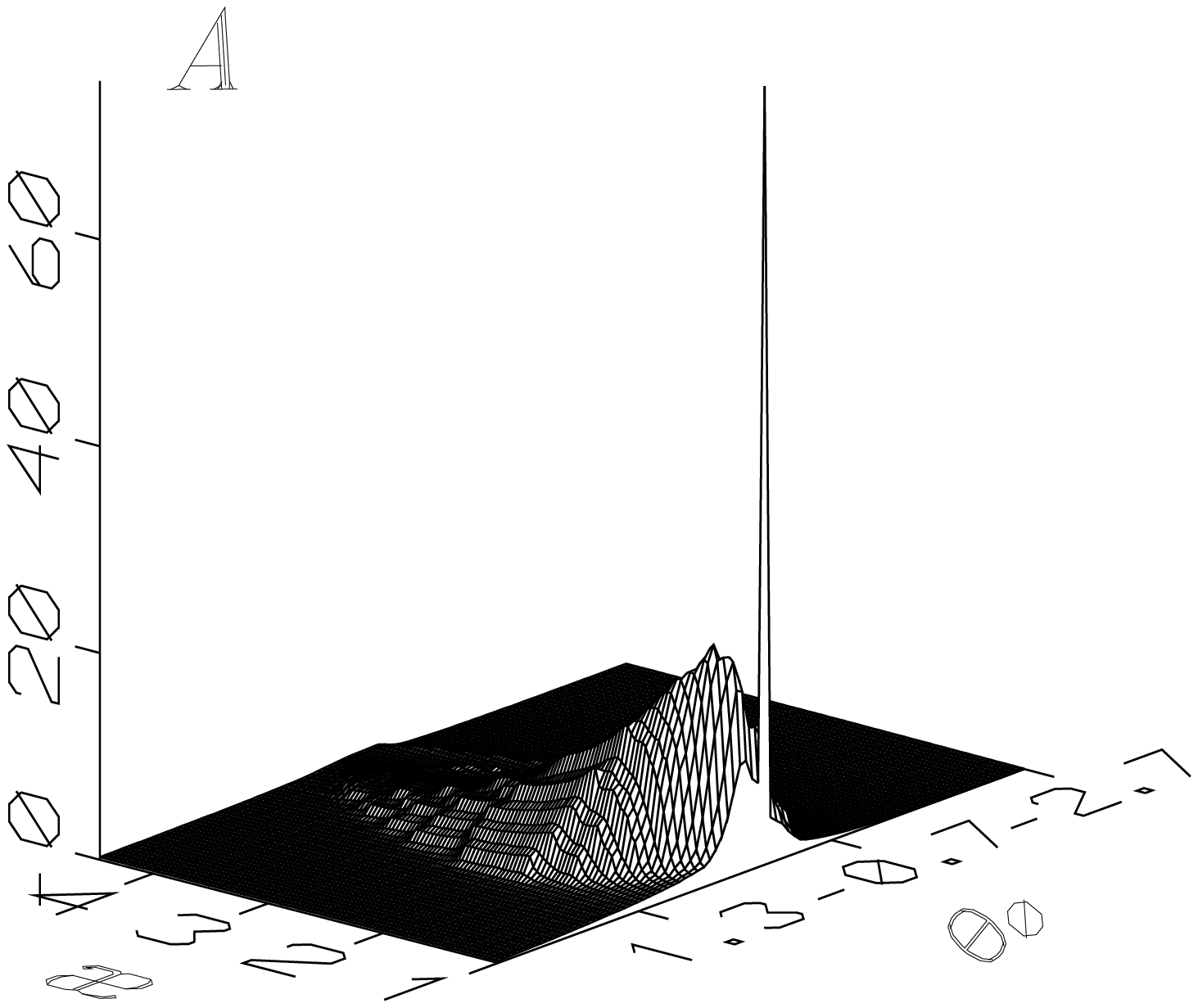}}

\end{center}

\begin{quote}
{\small Fig.3b. Dependence of resonance particle energy gain $A$ on 
perpendicular velocity $\,\ae_0\,$ and initial phase $\,\theta_0\,$.
The problem parameters are the following: $\;\chi =10^4,\;\nu =0,\;
\Lambda =10,\; \sigma =0.5,\; Y_0=4,\; u_0=u_s(Y_0)\,$ and 
$\; v_0=v_s(Y_0)\,$.}
\end{quote}
function $A$ dependence on $\ae_0$ 
when $\;\nu =0,\;\chi =10^4,\;\Lambda =10,\;\sigma =0.5\;$ and 
$\,Y_0=4\,$. The trapped particle initial conditions were 
$\;u_0=u_s(Y_0),\;\theta_0=\theta_b(\ae_0)\;$. The sharp peak
of $A(\ae )$ near the point $\ae =1$ corresponds to the 
synchronous particles population. According to Fig.3a, CRI-
efficiency for the tapped particles is less by factor of few
times than in the case of synchronous particles but it depends
more smoothly on the perpendicular velocity $\ae_0$ and the 
initial phase $\theta_0$. Therefore, the phase plane region occupied
by trapped particles is substantially larger than one occupied
by the synchronous particles. More evident it is demonstrated
in Fig.3b showing the efficiency $\, A(\ae_0 ,\theta_0 )\,$
dependence on both initial parmeters. Please pay attention to
high and narrow peak of $\, A(\ae_0 ,\theta_0 )\,$ corresponding 
to the synchronous particles and the low but wide maximum in
the case of trapped particles.
\vskip1.2cm

\begin{center}
{\small\bf 4. ANALYSIS OF WAVE FREQUENCY DRIFT INFLUENCE ON
RESONANT {\large\bf WPI} IN TRANSIENT PLASMA LAYER.}
\end{center}
\nopagebreak
\bigskip

Consider the wave frequency drift influence on the anomalous
cyclotron resonance interaction of energetic electron with a ducted 
whistler-mode wave propagating across the stationary transient
plasma layer. Suppose the magnetic field $Y(\xi )$ to be static
and given by (5) and the wave frequency sweeping to be described
by a function $h(p)$. Below $h(p)$ is taken as the linear function
$\, h(p)=p\,$. So the frequency sweeping rate becomes constant.
Contribution of the whistler frequency sweeping to the effective
potential well $\, U(\rho ,\theta )\,$ is characterized by a
function $r(\tau )$
$$
r=\frac{2}{Y^{1/2}Qv_s}\; \frac{d}{d\tau}\left[\, \frac{p}{Q^2}\;
(1+\frac{Y}{2})\,\right]\quad,
\eqno(10)
$$
$$
p\equiv\nu\left[\tau+\int\limits_{Y}^{Y_0}\frac{dY}{YF(Y)\; u_g(Y)}
\right]\; .
$$

Notice that for the parametrs values typical for the VLF-emission
generation in the magnetospher, the function $r(\tau )$ magnitude is 
usually small. The wave frequency drift causes the following 
modification of function $\rho $ determining the potential $U$
$$
\rho(\ae ,Y)=\ae g+(1+r-g)/\ae
\eqno(11)
$$

Consider the case when condition $\; 1+r-g>0\;$ is satisfied. Then
$\rho$ being a function of $\ae$, has minimum at $\; \ae_*(Y)=
[(1+r-g)\, /\,g]^{1/2}\;$ which is equal to $\;\min{\rho}\equiv\rho_*=
2[g\,(1+r-g)]^{1/2}\;$ and $\,\rho_*<1\,$ if the condition $\; r<
(2g-1)^2/\, 4g\;$ is fulfilled. So in the case $\; g-1<r<(2g-1)^2/\, 
4g\;$ there are trapped particles with perpendicular velocities $\ae$
in the range $\,\ae_1<\ae <\ae_2\,$, where $\ae_{1,2}$ are determined by
$$
\ae_{1,2}=\frac{1\pm\sqrt{(2g-1)^2-4rg}}{2g}\quad.
$$

On the phase plane $(\theta_{\tau},\theta)$ the phase trapping region has the 
maximum size at $\ae =\ae_*$ with its boundary defined by the following
equation
$$
0.5\; \theta_{\tau}^2+\Omega_b^2\, (\rho_*\theta -\sin{\theta})=
\Omega_b^2\, (\rho_*\arccos{\rho_*}-\sqrt{1-\rho_*^2})\quad.
$$

For faller with the frequency sweeping rate satisfying to the
condition $r<g\! -1\! <0\,$, $\rho(\ae )$ becomes zero at the point
$\,\ae\! =\!\ae_c(Y)\!\equiv\! [(g\! -\! 1\! -\! r)/g]^{1/2}\,$.
Therfore in this case there are always the trapped particles with 
the normalized perpendicular velocities $\ae$ in the range $\;\ae_3<
\ae <\ae_4\;$, where
$$
\ae_{3,4}=\frac{\sqrt{(2g-1)^2-4rg}\,\mp\, 1}{2g}\quad.
$$

Consider the whistler frequency drift influence on the cyclotron
resonance interaction of synchronous particle with initial data
$\, u_0=u_s(Y_0),\;\ae_0=1,\;\theta_0=0\,$. Introduce the specific
phase value $\;\theta_{\nu}=(2|r|)^{1/2}\;$ defined by the wave
frequency sweeping rate.

In the case of riser $\,r_0>0\,$ with the frequency sweeping rate
large enough the typical duration of synchronous particle CRI with
the whistler may be estimated as $\;\Delta\tau_R\simeq 4C_3\, /
\,(\chi |r|)^{1/2}\;$, where $C_3$ is a constant of the order of
unity. As a condition to suppress the anomalous CRI in TPL we take
the following: $\,\Delta\tau_R<1/4\,$. Hence we obtain the restriction
from below on the $r$-magnitude $\; r>r_c=6.5\cdot 10^4C_3^4/\chi^2\;$
and taking into account (5) it can be rewritten as a condition on
a parameter $\nu$: $\nu >\nu_c$, where $\nu_c$ is the critical value
of $\nu$. Consequently, in the case of riser with the frequency
sweeping rate large enough $(\nu >\nu_c)$, substantial decreasing of
the efficiency of synchronous particle cyclotron resonance interaction 
in TPL is occuring. In the range of initial phases $\, |\theta_0|<
\theta_{\nu}\,$, the energetic efficiency of CRI has the following scaling
on parameter $\nu$: $\; A\sim 1\, /\, (\chi^2|r|)^{1/4}\,$. If
$\, |\theta_0|>	\theta_{\nu}\,$, then this scaling is changed to one
given in Section $3$: $\, A\sim 1\, /\, (\chi |\theta_0|)^{1/2}\,$.

In the case of faller $(r<0)$, the synchronous particle becomes
trapped one. Therefore, under the same but small initial phases $\theta_0$
the duration of synchronous particle CRI for faller is substantially
larger than one for riser. The range of synchronous particles perpendicular
velocities is rather narrow $\,\delta\ae\equiv (\ae -1)\ll 1\,$, so
under the low frequency sweeping rate when $|r|\ll 1$, the potential
well $U$ is shrinking $\, U(\theta )\approx \Omega_b^2\, \{\theta\, [r+
(2g-1)\,\delta\ae\, ]+\theta^3/ 6\}\,$. Introduce notations $\, d\equiv
|r+(2g-1)\,\delta\ae\, |\,$ and $\,\theta_d=(2d)^{1/2}\ll 1\,$.
If $\,\chi\theta_d\gg 1\,$ the energetic efficiency can be scaled 
in the following way. When $\, |\theta_0|<\theta_d\,$ we obtain
$\, A\sim 1\,/\, (\chi\theta_d)^{1/2}\,$. In the range $\, |\theta_0|>
\theta_d\,$ the energetic efficiency is determined by the initial
phase detuning $\, A\sim 1\,/\, (\chi |\theta_0|)^{1/2}\,$.

The computer simulations of the cyclotron resonance interaction
efficiency under the whistler  mode wave propagating across the TPL
in the framework of equations (3) with $h(p)=p$ were performed 
with taking into account the wave frequency sweeping. For $\,\sigma =
0.5,\; \Lambda =10,\; \chi =10^4,\; Y_0=4,\; \theta_0=0,\; u_0=u_s(Y_0)\,$
and $\, v_0=v_s(Y_0)\,$, the results are shown in Fig.4 and Fig.5.
Fig.4a and Fig.5a depict efficiencies $A$ and $B$ in the case of
riser $\, (0<\nu <10^3\,)\,$. Fig.4b and Fig.5b are displaying them in the
case of faller when $\, 0<-\nu <10^{-2}\,$. For clarifying the 
efficiencies behaviour, the normalized variable $\chi |\nu |^{1/2}$
is used for the horisontal axis and plots of $\log A$, $\log B$ versus
$\, \log{(\chi |\nu |^{1/2})}\,$ are presented. 
According to Fig.4 and 
in correspondence with above developed theory function $A$ has the
plateau with the maximum value $A_m\approx 75$ in the range of small 
$|\nu |$ where the wave frequency drift influence is negligible.
Outside this region the energetic efficiency $A(\nu )$ falls
approximately as $\, A\sim 1\, /\, (\chi^2|\nu |)^{1/4}\,$ under $|\nu |$-
growth. For the positive $\nu$ (riser) the break of plot takes place 
at the point $(\,\chi^2|\nu |)^{1/2}\approx 2\,$.
\begin{center}
\vspace{0.5cm}

\mbox{\epsfysize=2in \epsffile{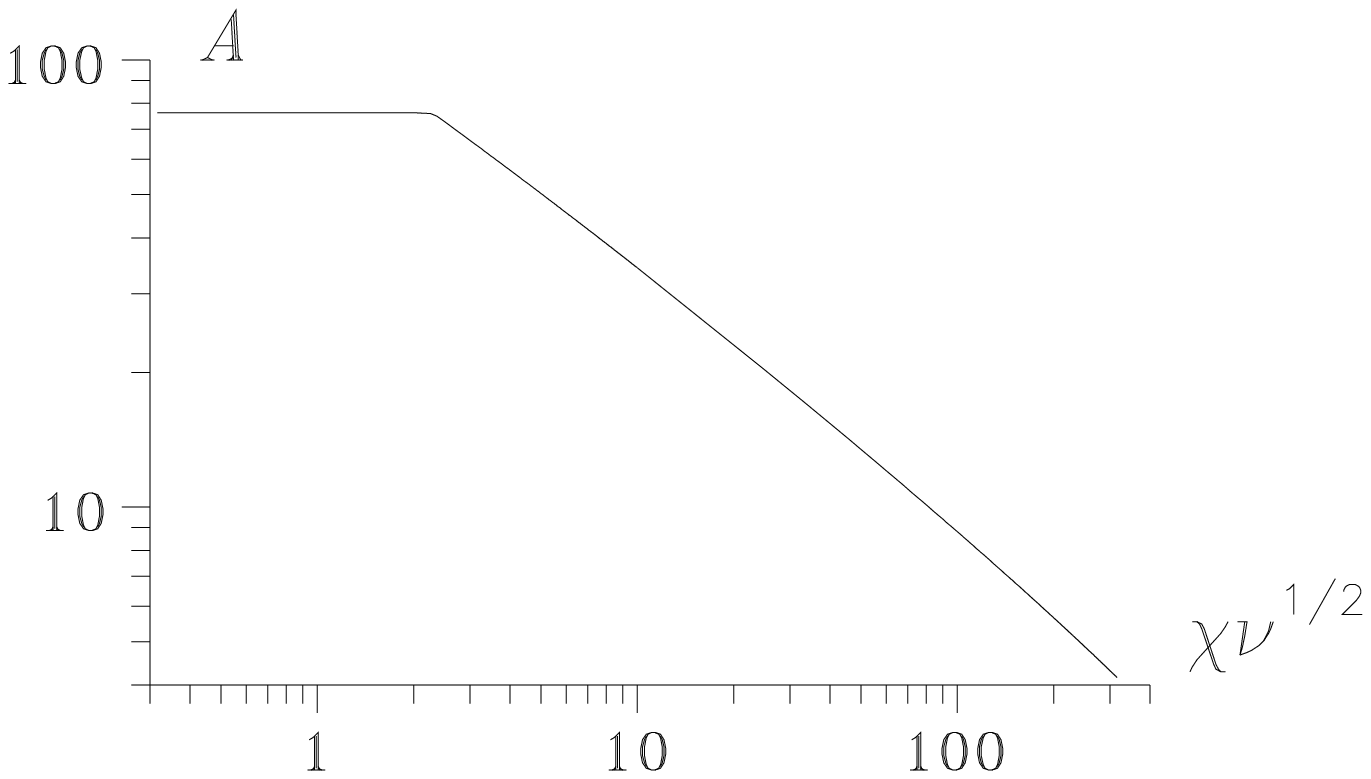}}

\end{center}
\begin{quote}
{\small Fig.4a. Dependence of synchronous particle energy gain $A$ on 
parameter $\,\chi\nu^{1/2}\,$ in the case of riser and $\;\chi =10^4$.}
\end{quote}
\begin{center}
\vspace{0.5cm}

\mbox{\epsfysize=2in \epsffile{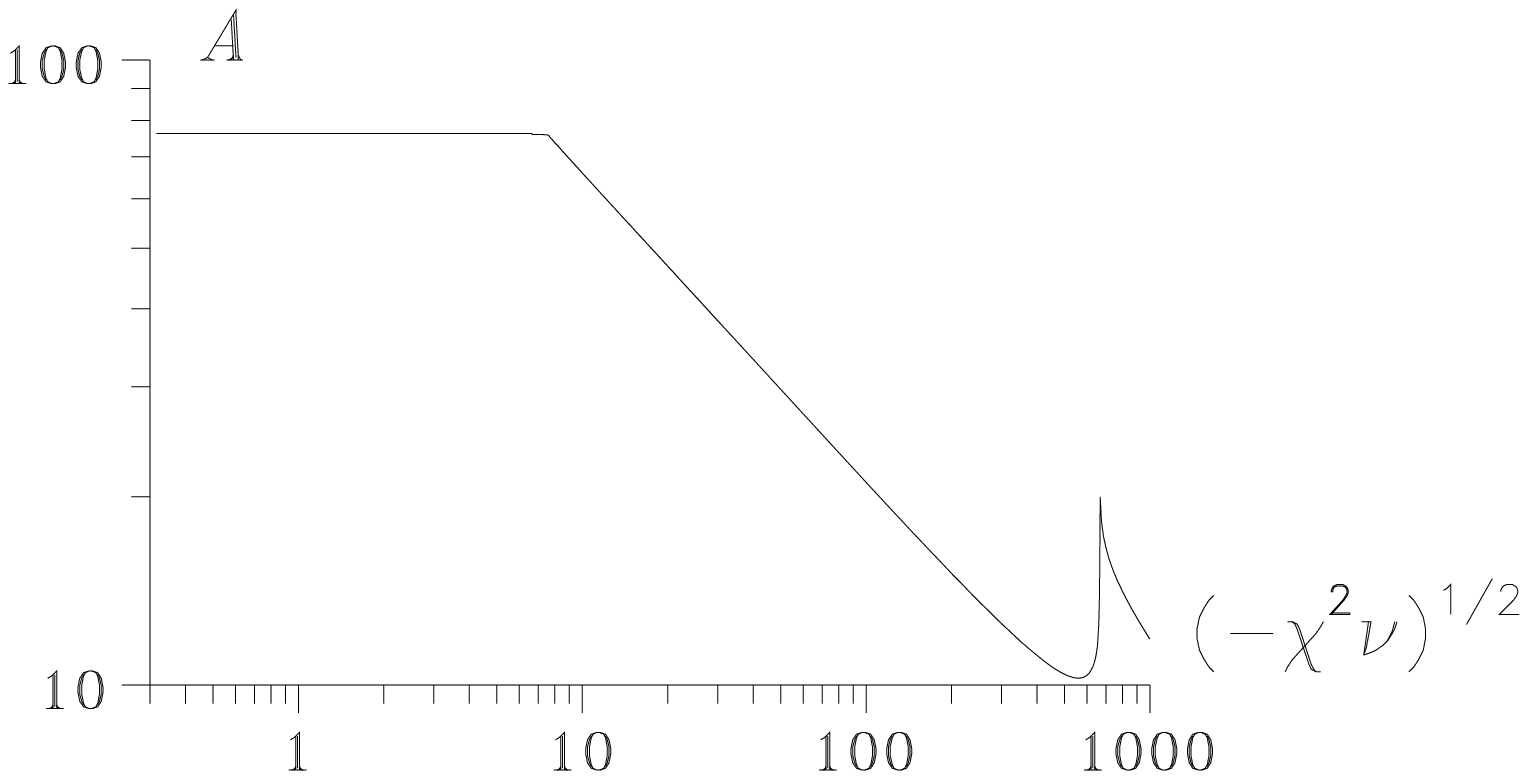}}

\end{center}
\begin{quote}
{\small Fig.4b. Dependence of synchronous particle energy gain $A$ on 
parameter $\,\chi |\nu|^{1/2}\,$ in the case of faller and $\;\chi =10^4$.}
\end{quote}

If $\nu $ is negative the plateau is substantially wider $\, (0<
(\chi^2|\nu |)^{1/2}\le 7)\,$. Besides at the point $(\,\chi^2|\nu |)^
{1/2}\approx 700\,$ the local maximums of $A$ and $B$ arise with 
$A\approx 16$ and $B\approx 51$ (see Fig.5b). Comparing plots of
functions $A$ and $B$ one can see that the pitch angle scattering
efficiency $B$ is substanially larger of the energetic efficiency $A$,
in particular, in the plateau region one has $B\approx 100$. Notice
also that outside the plateau region the pitch angle scattering
efficiency $B$ isn't governed by the simple power-law decay on
the variable $\, \chi |\nu|^{1/2}\,$

The numerical calculations of the trapped particles dependence
on the normalized perpendicular velocity $\ae$ and the wave
frequency sweeping parameter $\nu$ were made. Simulations results
are given in Fig.6 in the case of $\,\sigma =0.5,\;\Lambda =10,
\; Y_0=4\;$ and $\chi =10^4$. The initial data of resonance particle
correspond to the trapped electron located closely to the potential
well bottom, {\it i.e.} $\,u_0=u_s(Y_0),\; v_0=\ae\,v_s(Y_0)\,$
and $\; \theta_0=\cos^{-1}(\rho (\ae ,Y_0))\;$. Consider the 
trapped particle energy gain  in the case of its cyclotron resonance
interaction with riser (see Fig.6a). Notice the characteristic 
features of the normalized energy gain $A$. At first, there is
the sharp peak for small $\nu$ and $\,\delta\ae\equiv \ae -1\,$.
Secondly, outside this peak with parameter $\nu$ fixed, the energy 
gain $A$ has maximum on the variable $\ae$ at some point $\ae_m(\nu )$
corresponding to condition $\rho\approx 1$. In the range $\, 1<\ae
<\ae_m\,$ the trapped 
\begin{center}
\vspace{0.5cm}

\mbox{\epsfysize=2in \epsffile{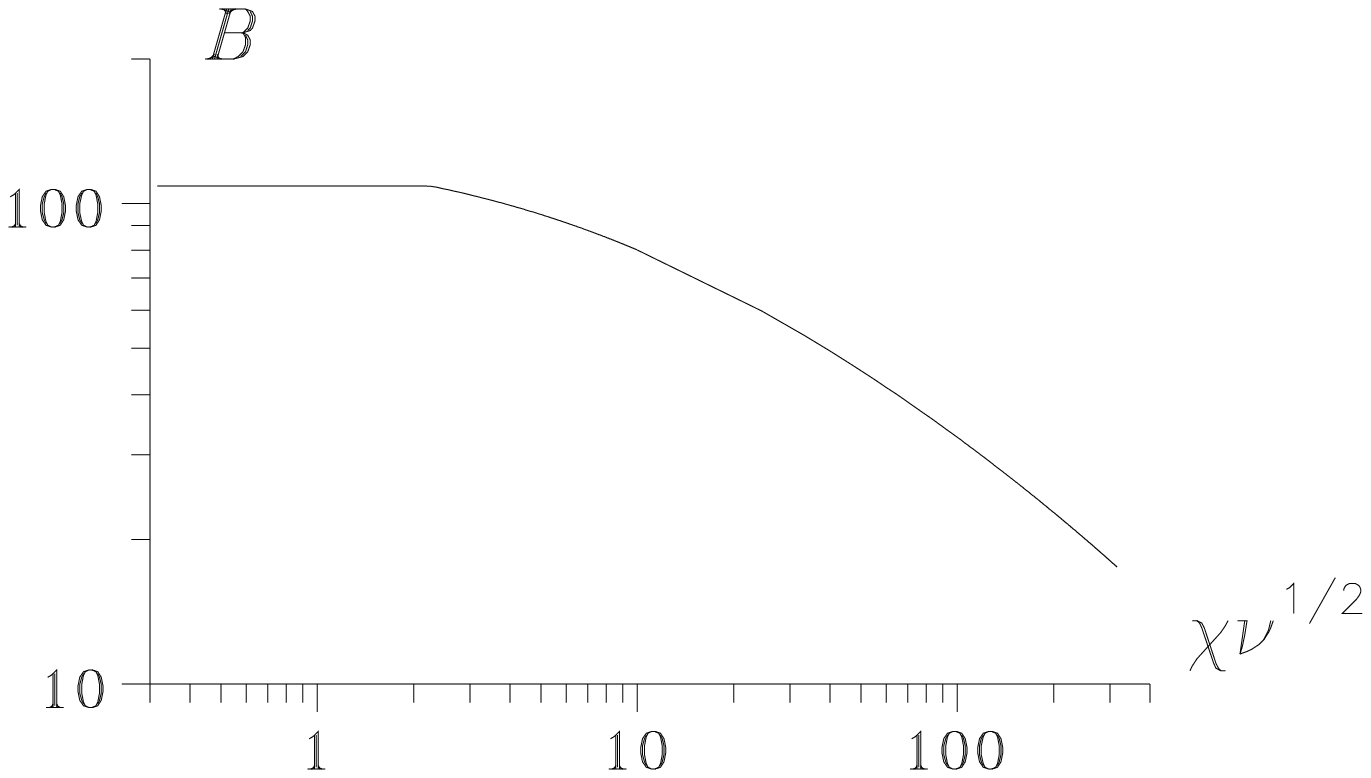}}

\end{center}
\bigskip

\begin{center}
\vspace{0.5cm}

\mbox{\epsfysize=2in \epsffile{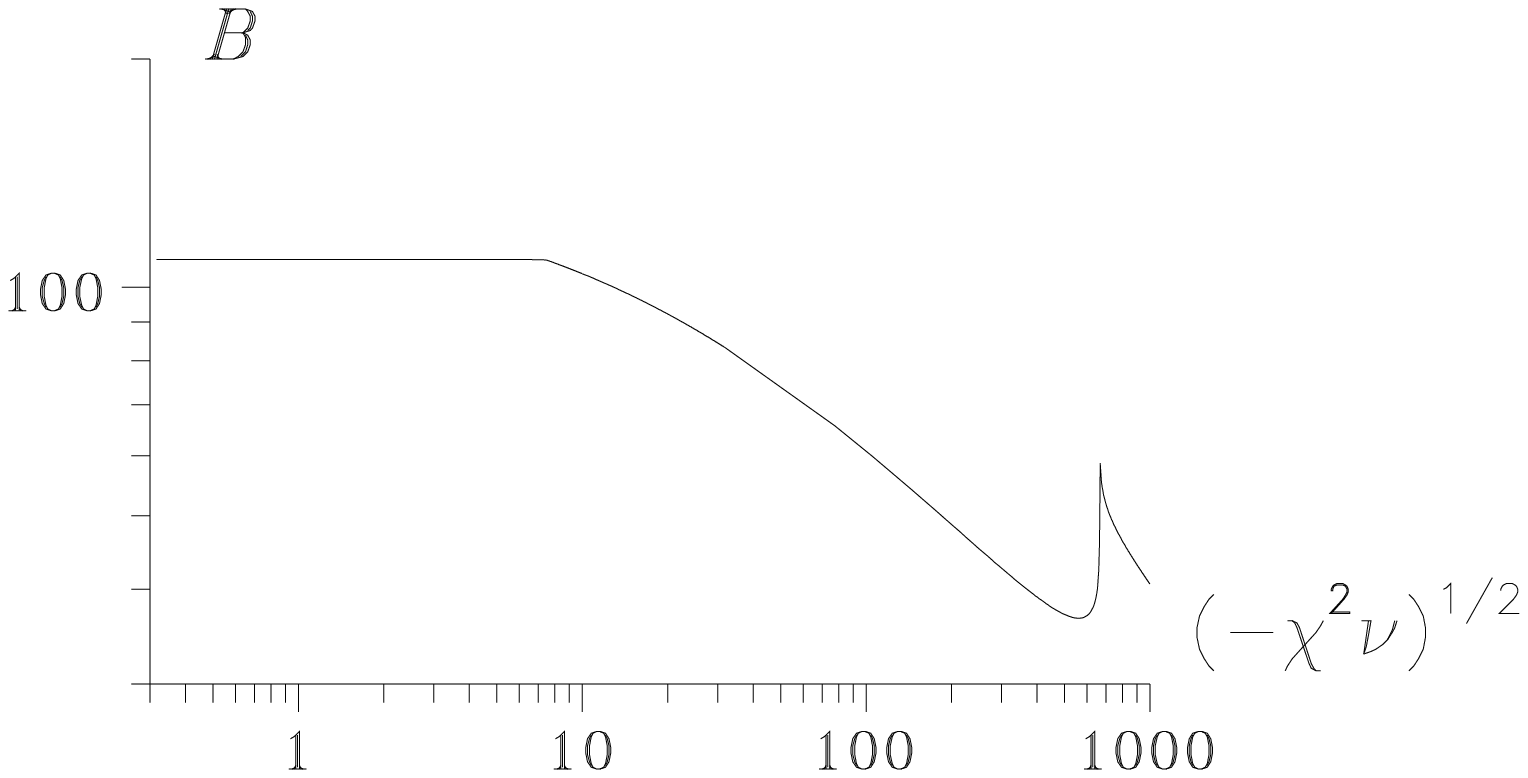}}

\end{center}

\begin{quote}
{\small Fig.5 Dependence of synchronous particle pitch angle scattering
efficiency $B$ on parameter $\,\chi |\nu|^{1/2}\,$: a - riser ($\nu >0$);
b - faller ($\nu <0$).}
\end{quote}
particles are absent so the energy gain is
very small. For $\ae >\ae_m$ the trapped particles exist and as
$\ae$ growth the energy gain $A$ is smoothly decaying.

The energy gain of trapped particles in the case of their interaction
with faller $(\nu >0)$ is plotted on Fig.6b. It can be seen that in
contrast to the riser case now the function $A(\ae ,\nu )$ does not
go down at the low $\delta\ae$. Thus the strong cyclotron resonance
interaction of trapped particles with faller is observed in the more
wide range of parameter $\bar{\nu}\equiv 10^2\,\nu$ variation. For
example, according to Fig.6b in the case when $\ae_0=1$ and $\bar{\nu}
=-3$, the energy gain of trapped particle with the initial phase
detuning $\theta_0\approx 0.46$ is equal to $A\approx 65\%$. This
particle crosses TPL in time $\,\Delta\tau\simeq 1.49\,$ and the
relative change of its pitch angle is $B\simeq 101\%\,$. Notice that 
for the wave frequency sweeping rate corresponding to $\Gamma =
4\cdot 10^{-5}\,$ and the dimensionless whistler amolitude
$\,\Omega_*\equiv1\, /\,\chi =10^{-4}\,$ the normalized parameter
$\bar{\nu}$ is equal to $\bar{\nu}=-3.18\,$.
\vskip1.2cm

\centerline{{\small\bf 5. CONCLUSION}}
\nopagebreak
\bigskip

The principal conclusions of analysis performed above are the following:

1. In the stationary transient plasma layer (5) the cyclotron resonance
interaction of energetic electrons with the small amplitude ducted 
whistler mode wave of the fixed fre- 
\begin{center}
\vspace{0.5cm}

\mbox{\epsfysize=2.5in \epsffile{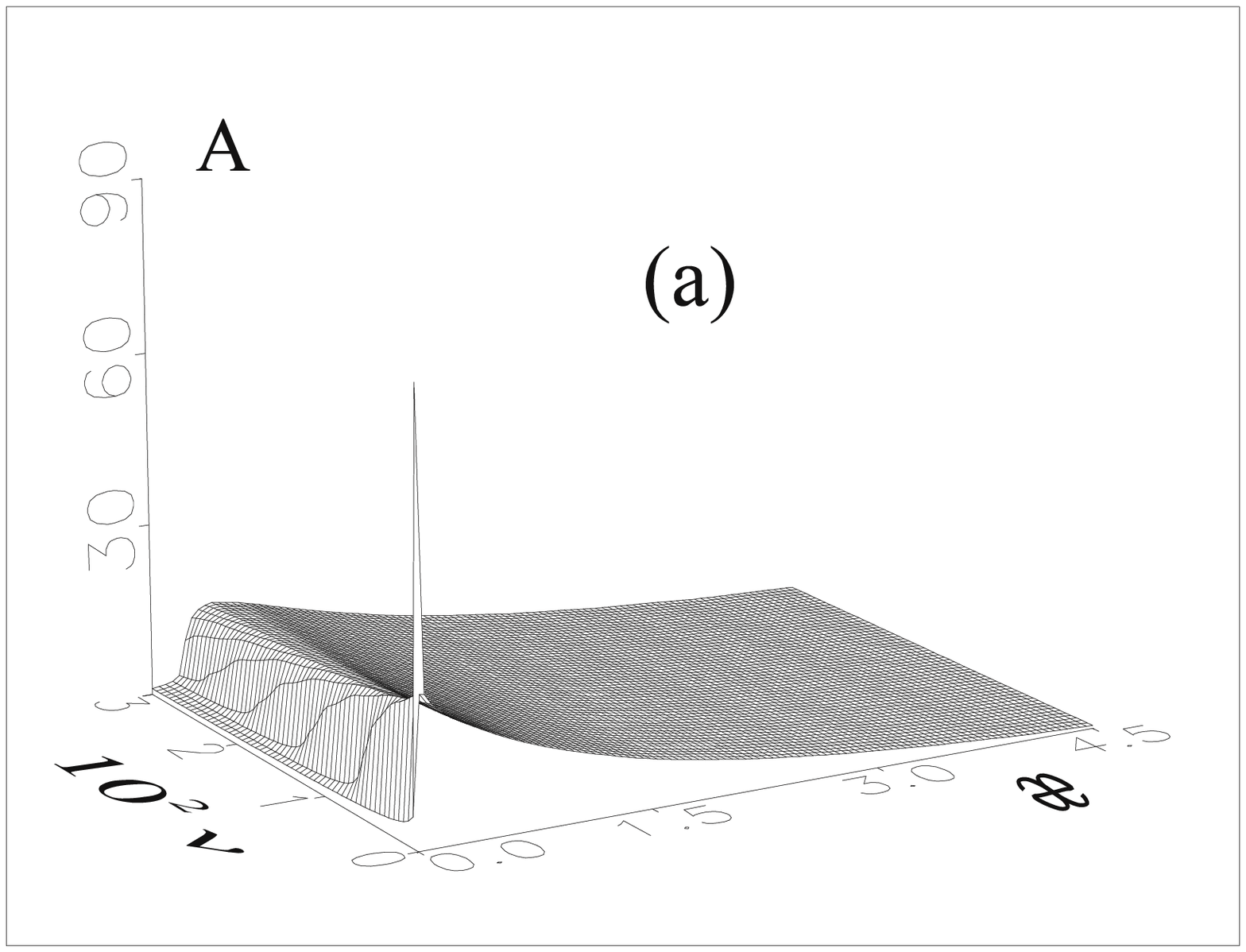}}

\end{center}

\begin{center}
\vspace{0.5cm}

\mbox{\epsfysize=2.5in \epsffile{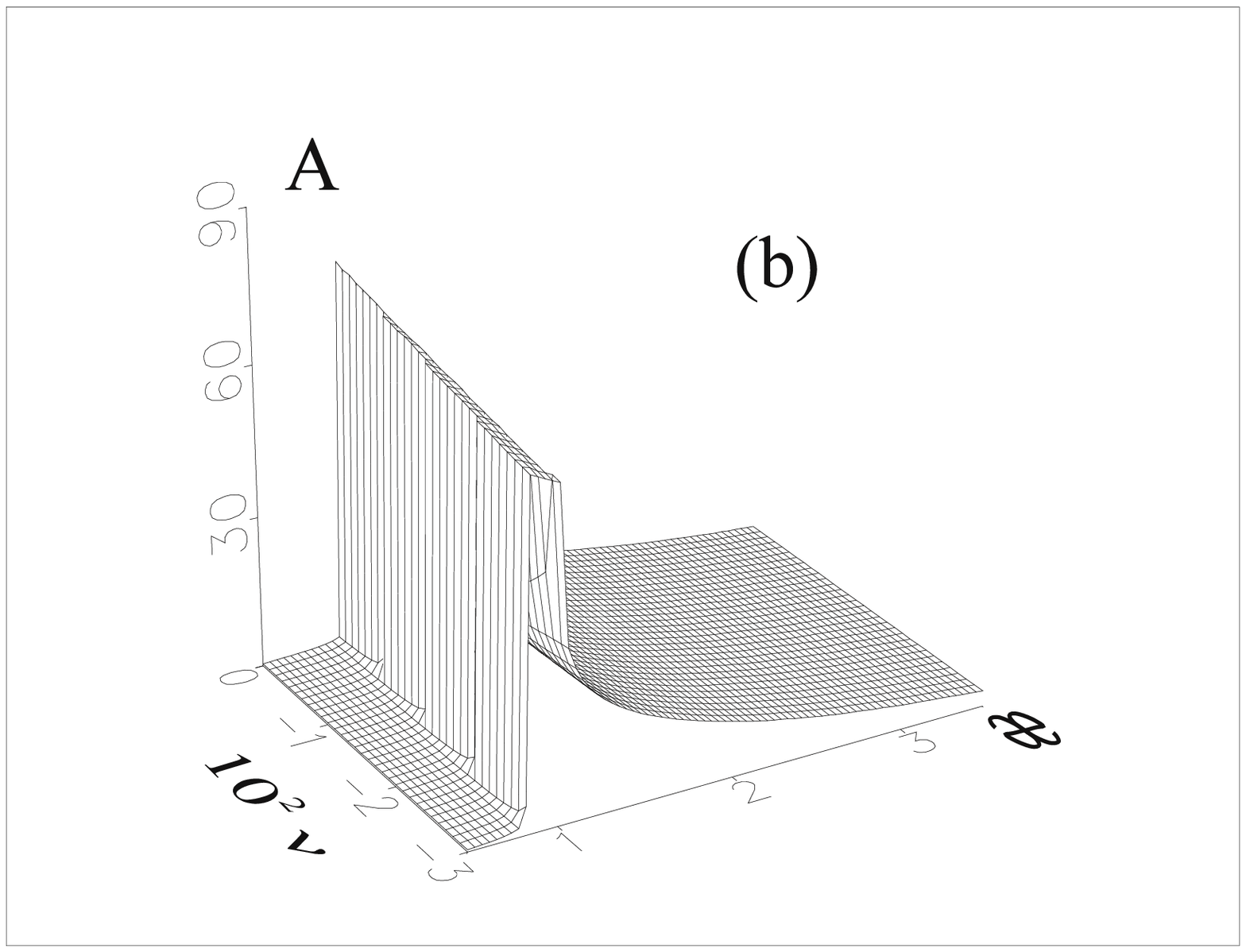}}

\end{center}

\bigskip
\begin{quote}
{\small Fig.6 Dependence of relative energy change of trapped particle
located close to potential well bottom on perpendicular velocity
$\,\ae\,$ and wave frequency sweeping parameter $\;\bar{\nu} =10^2\nu\,$:
a - riser ($\nu >0$); b - faller ($\nu <0$).}
\end{quote}
quency is determined by three
groups of resonance electrons: the synchronous particles, the trapped
and nontrapped resonance ones. For the synchronous particles the 
cyclotron resonance interaction is global in nature because the
resonanse region extends over the entire transient plasma layer.
If the magnetic field $B_0$ has the significant variation in TPL
{\it i.e.} $\delta B_0\sim B_0$, then the relative changes of
synchronous particle energy and pitch angle in course of the 
cyclotron resonance interaction are anomalously large up to the
order of $100\%$. For the trapped particles the CRI-duration
is less by a few times in comparison with the synchronous particles
one but this population is more numerous than the synchronous
particles group. Finally, in the case of nontrapped resonance particles
under the typical magnetospheric conditions the cyclotron resonance
interaction is weaker by the factor $10^2\div 10^3$ times in
comparison with the synchronous particles CRI. In turn on the phase 
plane the nontrapped resonance electrons population is the most
numerous. So it can be expected that in some range of the problem 
parameters all these groups of resonant particles may give comparable
contributions to the whistler damping (growth) rate in TPL.
However detailed calculations of the whistler damping (growth) rate
is a subject of a special consideration.

2. In the cyclotron resonance interaction of energetic electrons
with the whistler mode wave of variable frequency in the stationary
TPL, for synchronous particles with the initial data $\, u_0=u_s(Y_0),
\; v_0\approx v_s(Y_0)\,$ and $\Phi\approx\pi /2$ there is the 
critical value of wave frequency sweeping rate such that above it
$(\nu >\nu_0)$, the typical time scale of CRI decreases proportionally 
to $\, 1\, /\, (\chi^2|\nu |)^{1/4}\,$ where $\chi^2|\nu |>1$.
There is a plateau in the dependence of CRI-efficiency on parameter
$\chi^2|\nu |$ (see Fig.4 and Fig.5). The plateau width is larger
in the case of faller WPI.

3. In the case of riser with moderate frequency sweeping rate
corresponding to  $\bar{\nu}<1$, the anomalous CRI in the transient 
plasma layer takes place for the trapped particles with perpendicular
velocity in the range $\ae >1$. For the given $\nu$ the maximum
of energetic efficiency $A$ is observed at $\ae$ suiting to the condition
of the potential well being shrinked $(\rho(\ae ,\nu )\approx 1)$
and function $A$ decreases smoothly under $\ae$-growth.

4. In the case of energetic particles CRI with a faller in TPL, the
trapped particles anomalous CRI region on the perpendicular velocity
axis extends to the range $\ae <1$, where the maximum CRI-efficiency is
observed and it corresponds to $\rho (\ae ,\nu )\approx 1$.
Under parameters range considered in this paper the efficiency
of trapped particles CRI with a faller is larger approximately
two times than one in the case of riser.

In addition it is nesessary to note the following. Above we studied
the cyclotron resonance interaction of energetic electrons with the
whistler mode wave of variable frequency, propagating across the
stationary transient plasma layer. It was proved that there are the
synchronous particles groups whose CRI in TPL is of the global 
nature if the whistler frequency sweeping rate is low enough.
In the opposite case the synchronous particles existence may be
related to the following factors: a) the magnetic field $B_0$ 
nonstationarity; b) the specific modulation of whistler frequency
in analogy with one considered in paper of {\sc Brinca} (1981) and in 
paper of {\sc Bell} and {\sc Inan} (1981). In both cases a) and b)
there is a question about the temporal duration of ACRI under
a continuous injection of energetic particles into TPL. Besides, 
in the case a) it is also necessary to consider the source
of magnetic field $B_0(s,\tau )$ nonstationarity and to take
into account the influence of electric field $E_0$ driven by the time-
variable magnetic field $B_0$.
\vskip1.2cm

{\small{\bf Acknowledgments}-- We would like to thank Yu.V. Trakhtengerts
for helpful discussions concerning the influence of wave frequency
sweeping on ACRI.}
\bigskip

This work was supported in part by grant INTAS  94--2753 and --3120.

\vskip1.2cm

\centerline{\small\bf REFERENCES}
\nopagebreak
\bigskip
\paper{{\sc Bell T.F.} and {\sc Inan U.}}{1981}{{\small Transient nonlinear
pitch-angle scattering of energetic electrons by coherent VLF wave
packets in the magnetosphere. {\it J. Geophys. Res.}, {\bf 96}, 9047}} 

\paper{{\sc Bell T.F.}}{1984}{{\small The nonlinear gyroresonance
interaction between energetic electrons and coherent VLF waves
propagating at the arbitrary angle with respect to the Earth's magnetic
field. {\it J. Geophys. Res.}, {\bf 89}, 905}}

\paper{{\sc Brinca A.L.}}{1984}{{\small Enchancing whistler wave-electrons
interaction by the use of specially modulated injection.} {\it J. Geophys. 
Res.}, {\bf 86}, 702.}

\paper{{\sc Dowden R.L., McKay~A.D., Amon~L.E., Koens~H.C.} and 
{\sc M.H.~Dazey}}{1978}{{\small Linear and nonlinear amplification in the
magnetosphere during a 6.6-kHz transmission.} {\it J. Geophys. Res.}, 
{\bf 83}, 169.}

\paper{{\sc Dysthe K.B.}}{1971}{{Some studies of triggered whistler
emissions .} {\it J. Geophys. Res.}, {\bf 76}, 6915.}

\paper{{\sc Erokhin N.S.}}{1995}{{\small Long-term resonant wave-particle
interaction in inhomogeneous plasma. {\it In:} Second Volga international
summer school on space plasma physics, NIRFI, Nizhny Novgorod, p.33}}

\paper{{\sc Erokhin N.S. Michailovskaya L.A.} and {\sc N.N.~Zolnikova}}{1996}
{{\small Gyroresonant interaction of energetic electrons with whistler
mode wave in the transient layers of the circumterrestrial plasma.} 
{\it Geomag. and Aeronomy}, {\bf 36}, No 1.}

\paper{{\sc Helliwell R.A.}}{1967}{{\small A theory of discrete VLF
emissions from the magnetosphere.} {\it J. Geophys. Res.}, {\bf 72}, 4773.}

\paper{{\sc Helliwell R.A.}}{1993}{{\small 40 years of whistlers.
{\it In:} Modern Radio Sciences, Oxford Unversity Press, Oxford, p.189}}

\paper{{\sc Karpman V.I., Istomin J.N.} and {\sc D.R. Shklyar}}{1974}
{{\small Nonlinear theory of a quasi-monochromatic whistler mode
packet in inhomogeneous plasma.} {\it Plasma Physics}, {\bf 16}, 685}

\paper{{\sc Karpman V.I.} and {\sc D.R. Shklyar}}{1977} {{\small Particle
precipitation caused by a single whistler mode wave injection into 
the magnetosphere}. {\it Planet. Space Sci.,} {\bf 25}, 395}

\paper{{\sc Matsumoto H.}}{1979}{{\small Nonlinear whistler mode interaction
and triggered emissions in the magnetosphere: A review.{\it In:} Wave
Instabilities in space plasma, Palmadesso P.J. and K.Papadopoulos (eds),
p.163, D.Reidel, Dordrecht.}}

\paper{{\sc Molchanov O.A.}}{1985}{{\small Low frequency waves and
induced emissions radiation in the magnetosphere of the Earth, chapter 1,
Nauka, Moscow.}}

\paper{{\sc Nunn D.}}{1971}{{\small Wave-particle interactions in
electrostatic waves in an inhomogeneous medium.} {\it J. Plasma Physics},
{\bf 6}, 291.}

\paper{{\sc Nunn D.}}{1974}{{\small A self-consistent theory of triggered
VLF-emissions.} {\it Planet. Space Sci.}, {\bf 22}, 349.}

\paper{{\sc Nunn D.}}{1984}{{\small A quasistatic theory of triggered
VLF-emissions.} {\it Planet. Space Sci.}, {\bf 32}, 325.}

\paper{{\sc Omura Y., Nunn D., Matsumo- to H.} and {\sc M.J. Rycroft}}{1991}
{{\small A review of observational, theoretical and numerical studies
of VLF triggered emissions.} {\it Journ. Atmosph. Terrestr. Physics},
{\bf 53}, 351.}

\paper{{\sc Rycroft M.J.}}{1993}{{\small A review of whistler and
energetic electrons precipitation. {\it In:} Review of Radio Science
1990-1992, Oxford Science Publications, Oxford, p.631}}

\end{document}